\documentclass[%
 reprint,
 amsmath,amssymb,
 aps,
 prx,
 floatfix,
]{revtex4-2}
\usepackage{nicematrix}
\usepackage{mathtools}
\usepackage{graphicx}
\usepackage{dcolumn}
\usepackage{xcolor}
\usepackage{bm}

\usepackage[utf8]{inputenc} 
\usepackage{url}
\usepackage{amsmath}
\usepackage{amssymb}    
\usepackage{graphicx}
\usepackage{listings}
\usepackage{hyperref}

\renewcommand{\v}[1]{\ensuremath{\mathbf{#1}}} 
\newcommand{\gv}[1]{\ensuremath{\mbox{\boldmath$ #1 $}}} 
\newcommand{\grad}[1]{\nabla #1} 
\renewcommand{\div}[1]{\nabla \cdot #1} 
\newcommand{\pd}[2]{\frac{\partial #1}{\partial #2}} 
 
\renewcommand{\d}[2]{\frac{d #1}{d #2}} 
\newcommand{\delpar}[1]{\nabla_{\parallel} #1} 
\newcommand{\gradperp}[1]{\grad{}_{\perp} {#1}} 

\newcommand{\curv}[1]{{C}_{\left({#1}\right)}}

\newcommand{\cs}{c_s}

\newcommand{\bhatZ}{\v{b_0}}

\newcommand{\cur}{j_{\parallel}}

\newcommand{\vpi}{v_{\parallel i}}						
\newcommand{\vpe}{v_{\parallel e}}					
\newcommand{\gvort}{\omega}		

\newcommand{\momSrce}{S_{\mathcal{M}\parallel e}}							
\newcommand{\momSrci}{S_{\mathcal{M}\parallel i}}							
\newcommand{\nSrcN}{S_n}									
\newcommand{\enerSrceN}{S_{E,e}}							
\newcommand{\enerSrciN}{S_{E,i}}							

\newcommand{\alphad}{\alpha_d}

\newcommand{\epa}{\epsilon_R}
\newcommand{\epv}{\epsilon_v}
\newcommand{\epg}{\epsilon_G}
\newcommand{\epge}{\epsilon_{Ge}}

\newcommand{\kappai}{\kappa^i}
\newcommand{\kappae}{\kappa^e}
\newcommand{\BRef}{B_0}

\newcommand{\tRef}{t_0}

\newcommand{\nRef}{n_0}

\newcommand{\TeRef}{T_{e0}}
\newcommand{\TiRef}{T_{i0}}

\newenvironment{eqnal}{\equation\aligned}{\endaligned\endequation}

\newcommand{\fortranl}[1]{\lstinline[language={[90]Fortran}]{#1}}

\begin{document}

\preprint{APS/123-QED}

\title{Uncovering turbulent plasma dynamics via deep learning from partial observations}

\author{A. Mathews$^1$} \email{mathewsa@mit.edu} \author{M. Francisquez$^{1,2}$} \author{J.W. Hughes$^1$} \author{D.R. Hatch$^3$} \author{B. Zhu$^4$} \author{B.N. Rogers$^5$}
\affiliation{%
$^1$MIT Plasma Science and Fusion Center, Cambridge, Massachusetts 02139, USA
}%
\affiliation{$^2$Princeton Plasma Physics Laboratory, Princeton, New Jersey 08540, USA}
\affiliation{$^3$Institute for Fusion Studies, University of Texas, Austin, Texas 78704, USA}
\affiliation{$^4$Lawrence Livermore National Laboratory,  Livermore, California 94550, USA}
\affiliation{$^5$Department of Physics and Astronomy, Dartmouth College, Hanover, New Hampshire 03755, USA}

\date{\today}

\begin{abstract}

One of the most intensely studied aspects of magnetic confinement fusion is edge plasma turbulence which is critical to reactor performance and operation. Drift-reduced Braginskii two-fluid theory has for decades been widely applied to model boundary plasmas with varying success. Towards better understanding edge turbulence in both theory and experiment, we demonstrate that physics-informed neural networks constrained by partial differential equations can accurately learn turbulent fields consistent with the two-fluid theory from just partial observations of a synthetic plasma’s electron density and temperature in contrast with conventional equilibrium models. These techniques present a novel paradigm for the advanced design of plasma diagnostics and validation of magnetized plasma turbulence theories in challenging thermonuclear environments. 

\end{abstract}

\maketitle


\section{\label{sec:level1} Introduction}

Predicting turbulent transport in the edge of magnetic confinement fusion devices is a longstanding goal spanning several decades currently presenting significant uncertainties in the particle and energy confinement of fusion power plants \cite{Carreras_1997, Chapman2019UKAEACT}. The boundary region is critical in determining the fusion device's overall viability since edge plasma conditions strongly influence a myriad of reactor operations ranging from core fuelling to power output to wall safety \cite{Federici_2001,Li2020,Chang2021}. Yet edge modelling continues to need improvement---comprehensive gyrokinetic codes suitable for the boundary of tokamaks are still under development and fluid simulations commonly lack essential physics necessary to study collisionless systems. One particular transport theory relevant to boundary plasmas and widely applied to analyze edge turbulence is the drift-reduced Braginskii model \cite{Braginskii1965RvPP,Kim-1993,francisquez_thesis}. Various adaptations of these equations have been recently taken to investigate several important edge phenomena including pedestal physics \cite{Francisquez_2017}, blob dynamics \cite{Ricci_2012}, neutral effects \cite{Thrysoe_2018}, and heat fluxes impinging plasma-facing components \cite{nespoli_non-linear_2017}. And while various trends are at times reproduced in these works, direct quantitative agreement between the two-fluid turbulence theory and observations is generally lacking on a wide scale due to difficulty in aligning global simulations with intricate plasma experiments where relevant measurements may be sparse or missing altogether. Yet this is a crucially important endeavour since gathering sufficient information to effectively test reduced turbulent transport models is vital to developing predictive capability for future fusion devices. These machines will access novel burning plasma regimes and operate with some of the largest temperature gradients in the universe, but our existing models may be inaccurate and standard diagnostics incapable of surviving such harsh thermonuclear environments \cite{biel_diagnostics_2019}. Fusion plasma diagnostic measurements accordingly require suitable analysis techniques as they are inherently noisy and limited in their spatiotemporal scope such as 1- or 2-dimensional profiles of electron density and temperature \cite{griener_continuous_2020,Bowman_2020,Mathews2020,Mathews2021}. Furthermore, it is essential to know the turbulence-generated edge electric field since ${\bf E \times B}$ drifts, which strongly affect perpendicular transport, influence plasma stability and structure across the edge \cite{Simakov-paper,osti_1630104, osti_1762124}. Resultant downstream interactions from turbulent particle and heat fluxes striking surfaces therein pose significant risk to safe operation, e.g. the control of radiative regimes such as detached divertor schemes, in upcoming fusion reactors \cite{kuang_SPARC,kuang_ARC}. To this end, we demonstrate a novel physics-informed deep learning technique capable of diagnosing unknown turbulent field fluctuations directly consistent with drift-reduced Braginskii theory from limited electron pressure observations. 

We represent the drift-reduced Braginskii model via physics-informed neural networks (PINNs) \cite{Lagaris_1998,raissi2017physics,MICHOSKI2020193,SIRIGNANO20181339}---highly expressive function approximators trained to solve supervised learning tasks while respecting nonlinear partial differential equations---to infer unobserved field dynamics from partial measurements of a synthetic plasma. As illustrated through a readily adaptable multi-network machine learning framework, this paradigm is transferable to the broad study of quasineutral plasmas in magnetized collisional environments and presents novel pathways for the AI-assisted design of plasma diagnostics. In ways previously inaccessible with classical analytic methods, this framework has the ability to improve the direct testing of reduced turbulence models in both experiment and simulation to inform discovery of the equations necessary to model the edge. The overall computational technique introduces significant advances in systematizing the quantitative validation of plasma turbulence theories and is to date among the most complex systems applied in physics-informed deep learning codes.


To demonstrate this framework, we proceed with a description of drift-reduced Braginskii theory and computational modelling of a synthetic plasma in Section \ref{sec:level2}, outline a novel physics-informed machine learning architecture suited for the analysis of multi-field plasma turbulence in Section \ref{sec:level3}, present results in the robust learning of unknown turbulent fields in section \ref{sec:level4}, and conclude with a summary and future outlook in Section \ref{sec:level5}.

\section{\label{sec:level2} Synthetic plasma construction}

The synthetic plasma analyzed is numerically simulated by the global drift-ballooning (GDB) finite difference code \cite{osti_1488775, francisquez2020fluid} which solves the two-fluid drift-reduced Braginskii equations in the electrostatic limit relevant to low-$\beta$ conditions. This is a full-$f$ \cite{Belli_2008,Full-F,Held_2020} fluid model in the sense that the evolution of the equilibrium and fluctuating components of the solution are not separated and relative perturbation amplitudes can be of order unity as found in experiments \cite{Garcia_SOL_fluctuations}. The plasma is magnetized and quasineutral with the perpendicular fluid velocity given by ${\bf E \times B}$, diamagnetic, and ion polarization drifts. After neglecting collisional drifts, as well as terms of order $m_e/m_i$, one arrives at the following set of equations (in Gaussian units) governing the evolution of the synthetic plasma's density ($n \approx n_e \approx n_i$), vorticity ($\gvort$), parallel electron velocity ($\vpe$), parallel ion velocity ($\vpi$), electron temperature ($T_e$), and ion temperature ($T_i$) \cite{francisquez2020fluid}

\begin{eqnal}\label{eq:nDotGDBH}
\d{^e n}{t} = -\frac{2c}{B}\left[n\curv{\phi}-\frac{1}{e}\curv{p_e}\right] -n\delpar{\vpe} +\nSrcN+\mathcal{D}_{n}
\end{eqnal}
\begin{eqnal}\label{eq:wDotGDBH}
\pd{\gvort}{t} &= \frac{2c}{eB}\left[\curv{p_e}+\curv{p_i}\right]-\frac{1}{em_i \Omega_i}\curv{G_i} \\
&\quad+\frac{1}{e}\delpar{\cur}-\div{\left\lbrace\frac{nc^2}{\Omega_i B^2}\left[\phi,\gradperp{\phi}+\frac{\gradperp{p_i}}{en}\right]\right. \\ 
&\quad\left.+\frac{nc}{\Omega_i B}\vpi\delpar{\left(\gradperp{\phi}+\frac{\gradperp{ p_i}}{en}\right)}\right\rbrace}+\mathcal{D}_{\gvort}
\end{eqnal}
\begin{eqnal}\label{eq:vpeDotGDBH}
\d{^e\vpe}{t} &= \frac{1}{m_e}\left(e\delpar{\phi}-\frac{\delpar{p_e}}{n}-0.71\delpar{T_e} + e\eta_\parallel\cur \right) \\
&\quad + \frac{2}{3} \frac{\delpar{G_e}}{n} + \frac{2cT_e}{eB}\curv{\vpe}+\momSrce+\mathcal{D}_{\vpe}
\end{eqnal}
\begin{eqnal}\label{eq:vpiDotGDBH}
\d{^i\vpi}{t} &= \frac{1}{m_i}\left(-e\delpar{\phi}-\frac{\delpar{p_i}}{n}+0.71\delpar{T_e} - e\eta_\parallel\cur \right)\\
&+\frac{2 T_e}{3n}\frac{\delpar{G_i}}{n}-\frac{2cT_i}{eB}\curv{\vpi}+\momSrci+\mathcal{D}_{\vpi}
\end{eqnal}
\begin{eqnal}\label{eq:TeDotGDBH}
\d{^e T_e}{t} = \frac{2T_e}{3n}\left[\d{^e n}{t} + \frac{1}{T_e}\delpar \kappa^e_\parallel \delpar T_e + \frac{5n}{m_e \Omega_e} \curv{T_e} \right.\\ \left. + \eta_\parallel \frac{\cur^2}{T_e} + \frac{0.71}{e}(\delpar{\cur} - \frac{\cur}{T_e}\delpar{T_e}) + \frac{1}{T_e} \enerSrceN \right] + \mathcal{D}_{T_e},
\end{eqnal}
\begin{eqnal}\label{eq:TiDotGDBH}
\d{^i T_i}{t} &= \frac{2T_i}{3n}\left[\d{^i n}{t} + \frac{1}{T_i}\delpar \kappa^i_\parallel \delpar T_i \right.\\
&\quad \left. - \frac{5n}{m_i \Omega_i} \curv{T_i} + \frac{1}{T_i} \enerSrciN \right] + \mathcal{D}_{T_i}
\end{eqnal}
whereby the field-aligned electric current density is $\cur = en\left(\vpi - \vpe\right)$, the stress tensor's gyroviscous terms contain $G_s = \eta^s_0 \left\lbrace 2\delpar{v_{\parallel s}}+c\left[\curv{\phi} + \curv{p_s}/(q_s n)\right]\right\rbrace$, and $\eta^s_0$, $\Omega_s$, and $q_s$ are the species ($s = \{e,i\}$) viscosity, cyclotron frequency, and electric charge, respectively. The convective derivatives are $d^s f/dt = \partial_t f + (c/B)\left[\phi,f\right] + v_{\parallel s}\delpar{f}$ with $\left[F,G\right] = \bhatZ \times \nabla F \cdot \nabla G$ and $\bhatZ$ representing the unit vector parallel to the magnetic field. The field's magnitude, $B$, decreases over the major radius of the torus ($B\propto1/R$), and its curvature is $\gv{\kappa} = -{\bf{\hat{R}}}/R$. The curvature operator, $\curv{f} = \bhatZ \times \gv{\kappa} \cdot \grad{f}$, $\nabla_\parallel = -\partial / \partial z$, and $\bhatZ = -{\bf \hat{z}}$ follow past convention \cite{francisquez2020fluid}. The coefficients $\kappa^s_\parallel$
and $\eta^s_\parallel$ correspond to parallel thermal conductivity and electrical resistivity, respectively. Time-independent Gaussian density ($\nSrcN$) and energy sources ($S_{E,s}$) are placed at the left wall while zero external momentum ($S_{\mathcal{M}\parallel s}$) is explicitly forced upon the system. Explicit hyperdiffusion consisting of both fourth-order cross-field and second-order parallel diffusion is applied for numerical stability in the form of $\mathcal{D}_f = \chi_x \frac{\partial f} {\partial x^4} + \chi_y \frac{\partial f} {\partial y^4} + \chi_z \frac{\partial f} {\partial z^2}$. Under quasineutrality, electric fields arise not by local imbalance of charged particles but by the requirement that the electric current density is divergence free \cite{DRB_consistency3,Zholobenko_2021}. Accordingly, the electrostatic potential, $\phi$, is numerically solved via the following boundary value problem: \begin{equation}\label{BVP_gvort_phi}
\div{ \frac{nc}{\Omega_i B}\left(\gradperp{\phi}+\frac{\gradperp{p_i}}{en}\right) } = \gvort.
\end{equation}

The 3-dimensional simulation domain implemented is a shearless field-aligned coordinate system where ${\bf \hat{x}}$ is the unit vector along the radial direction (i.e. ${\bf{\hat{R}}}$), the helical magnetic field is oriented along ${\bf \hat{z}}$, and ${\bf \hat{y}}$ is perpendicular to both ${\bf \hat{x}}$ and ${\bf \hat{z}}$. The synthetic plasma consists of deuterium ions and electrons with real masses (i.e. $m_i = 3.34 \times 10^{-27} \text{ kg}$ and $m_e = 9.11\times 10^{-31} \text{ kg}$) and on-axis magnetic field of $B_{axis} = 5.0 \text{ T}$ with minor and major radius of $a_0 = 0.22 \text{ m}$ and $R_0 = 0.68 \text{ m}$, respectively, consistent with characteristics of discharges in the high-field Alcator C-Mod tokamak \cite{Alcator_Greenwald} for which there is notable evidence of fluid drift turbulence controlling edge dynamics \cite{labombard_evidence_2005}. Moreover, drift-reduced fluid models, where the ion gyration frequency is considered to be faster than the evolving turbulent plasma fluctuations (i.e. $\Omega_i \gg \frac{\partial}{\partial t}$), are generally good approximations to full velocity models when studying edge turbulence \cite{Leddy_full_velocity}. 

This discretized toroidal geometry is a flux-tube-like domain on the outboard side (i.e. strictly bad-curvature) of the tokamak with field lines of constant helicity wrapping around the torus and terminating on walls producing both resistive interchange and toroidal drift-wave turbulence. Transport is primarily along blobby field-aligned structures with increased pressure propagating due to perpendicular drifts which polarize the blob and yield outward ${\bf E \times B}$ drift of the filament. This is related to the Poynting vector representing the directional energy flux density of the electromagnetic field \cite{Thrysoe_2020,DRB_consistency3}. The physical dimensions of the entire simulation domain are $[L_x = 7.7\text{ cm}, L_y = 5.5 \text{ cm}, L_z = 1800.0 \text{ cm}]$ with spatiotemporal resolution of $[\Delta x = 0.03 \text{ cm}, \Delta y = 0.04 \text{ cm}, \Delta z = 56.25 \text{ cm}, \Delta t = 4.55 \times 10^{-11} \text{ s}]$. Periodic boundary conditions are employed in the binormal direction for all quantities. Homogeneous Neumann conditions are set in the radial coordinate for $n$, $\vpe$, $\vpi$, $T_e$, and $T_i$ while homogeneous Dirichlet conditions are used for $\gvort$ and $\phi$. By constraining $\phi = 0$ along the walls, this in principal enforces radial $\bf {E \times B}$ flows to go to zero on the boundaries of the simulation domain. The lower limit of the Bohm criterion is imposed as a sheath condition,
\begin{equation}
\vpi(z=\pm \frac{L_z}{2}) = \mp c_{s} = -\sqrt{\frac{T_i + T_e}{m_i}},
\end{equation}
\begin{equation}
\vpe(z=\frac{\pm L_z}{2}) = \begin{cases} \mp c_{s}\exp(\Lambda - \frac{e\phi}{T_e}) &\mbox{if } \phi > 0 \\
\mp c_{s}\exp(\Lambda) &\mbox{if } \phi \leq 0 \end{cases},
\end{equation}
where $\Lambda = \log\sqrt{m_i/[2\pi m_e(1 + \frac{T_i}{T_e})]}$. Since the direction of the flows at the sheath entrance are known, ghost cells in the $z$-direction are filled such that an upwind stencil ensues to evolve $n$, $\gvort$, $T_e$, and $T_i$ \cite{francisquez2020fluid}. For $T_e$ and $T_i$ specifically, finite conductive heat fluxes entering the sheaths are applied according to $q_{{\parallel},s} = -\kappa^s_{\parallel} \delpar T_s = \pm \gamma_s n v_{{\parallel},s} T_s$,
where the upper (lower) sign corresponds to the top (bottom) sheath and $\gamma_s$ is the sheath transmission coefficient. Its value for ions and electrons is taken to be $\gamma_i =5T_i/2T_e$ and $\gamma_e = 2 + \lvert e \phi \rvert/T_e$, respectively \cite{francisquez2020fluid}. Collisional coefficients and diffusivities are kept constant in the direct numerical simulation as they can be unphysically large at high temperatures due to the lack of kinetic effects and generally require closures going beyond Chapman-Enskog. To start the numerical simulation, electrons and ions are initialized with zero parallel velocity and vorticity fields along with truncated Gaussian density and temperature profiles. A second-order trapezoidal leap-frog time-stepping scheme evolves the system of equations forward with subcycling of parabolic terms (e.g. $\delpar \kappa^s_\parallel \delpar T_s$) due to the low frequency turbulence structure changing slowly over the thermal diffusion timescale. The commonly applied Boussinesq approximation \cite{GRILLIX_2018} in Braginskii solvers is also used when evolving the generalized vorticity, $\gvort$. The normalizations applied to solve these partial differential equations in both the finite difference code and deep learning framework are sketched in the Appendix. A complete treatment of the numerical solver and greater specificity regarding the GDB code can be found in \cite{osti_1488775}.

\section{\label{sec:level3} Machine learning plasma theory}

Neural networks are operationally computational programs composed of elementary arithmetic operations (e.g. addition, multiplication) and functions (e.g. $\exp$, $\sin$, $\log$) which can be differentiated to arbitrary order up to machine precision via application of chain rule \cite{Raissi_JMLR,AD_2020}. While biases are presently inevitable \cite{wang2020eigenvector}, these regression models are in theory constructed without necessarily committing to a designated class of basis functions (e.g. polynomial, trigonometric). Automatic differentiation in conjunction with this adaptive capacity of neural networks permits them to effectively address nonlinear optimization problems in physics and engineering by training upon both partial differential equations and observational data via multi-task learning \cite{raissi2017physics}. Constraining classically underdetermined systems by physical laws and experimental measurements in this way presents an emerging technique in computational mechanics which we extend to the deduction of unknown turbulent plasma dynamics. In this deep learning framework, every dynamical variable in equations \eqref{eq:nDotGDBH}--\eqref{eq:TiDotGDBH} is approximated by its own fully-connected neural network, which is commonly known as a data-efficient universal function approximator \cite{cybenko_approximation_1989}, since its high plasticity enables it to be molded given sufficient training. 

\begin{figure}[ht]
\includegraphics[width=1.0\linewidth]{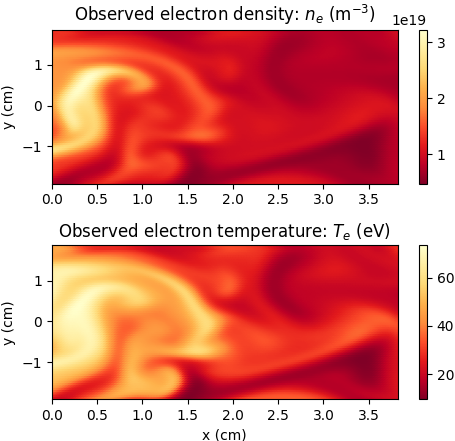}
\caption{\label{observed_dens_and_Te} These 2-dimensional measurements of turbulent electron density and temperature over a short temporal window are the only observed variables from the 3-dimensional synthetic plasma exhibiting blob-like filaments.}
\end{figure}

For analysis in the multi-network framework, partial measurements of $n_e$ and $T_e$ over time only come from a smaller 2-dimensional field-aligned domain in the interior of the synthetic plasma to emulate experiment (e.g. gas puff imaging \cite{Zweben_2017}) with dimensions of $[L^*_x = 3.8 \text{ cm}, L^*_y = 3.8 \text{ cm}]$ and spatiotemporal resolution of $[\Delta^* x = 0.03 \text{ cm}, \Delta^* y = 0.04 \text{ cm}, \Delta^* t = 7.27 \times 10^{-7} \text{ s}]$ as depicted by a snapshot in Figure \ref{observed_dens_and_Te}. Each network consequently takes the local spatiotemporal points $(x,y,t)$ from the reduced domain for measurements as the only inputs to the initial layer while the dynamical variable being represented is the sole output. In the middle of the architecture, every network consists of 5 hidden layers with 50 neurons per hidden layer and hyperbolic tangent activation functions ($\sigma$) using Xavier initialization \cite{GlorotAISTATS2010}. 

\begin{figure}[ht]
\includegraphics[width=1.0\linewidth]{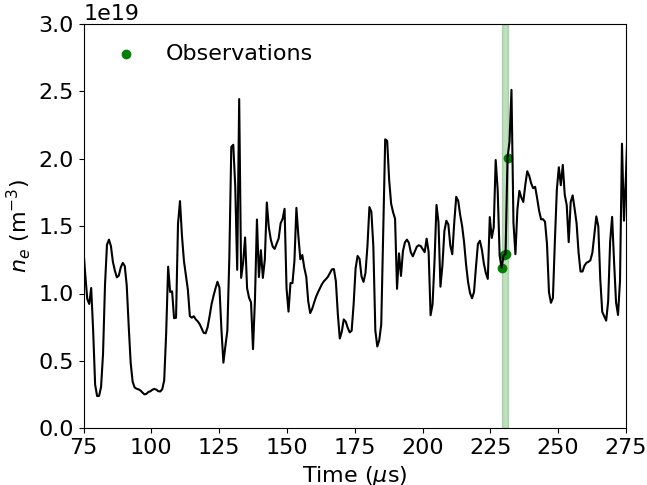}
\caption{\label{observed_1D_dens_time} A local time trace of the turbulent $n_e$ over 200 $\mu$s from the simulated plasma at $[x = 1.0 \text{ cm}, y = 0.0 \text{ cm}, z = -28.1 \text{ cm}]$. The observed synthetic data analyzed in the machine learning framework only comes from the small temporal window (green) which corresponds to just 4 points in time.}
\end{figure}

\begin{figure*}
\includegraphics[width=0.8\linewidth]{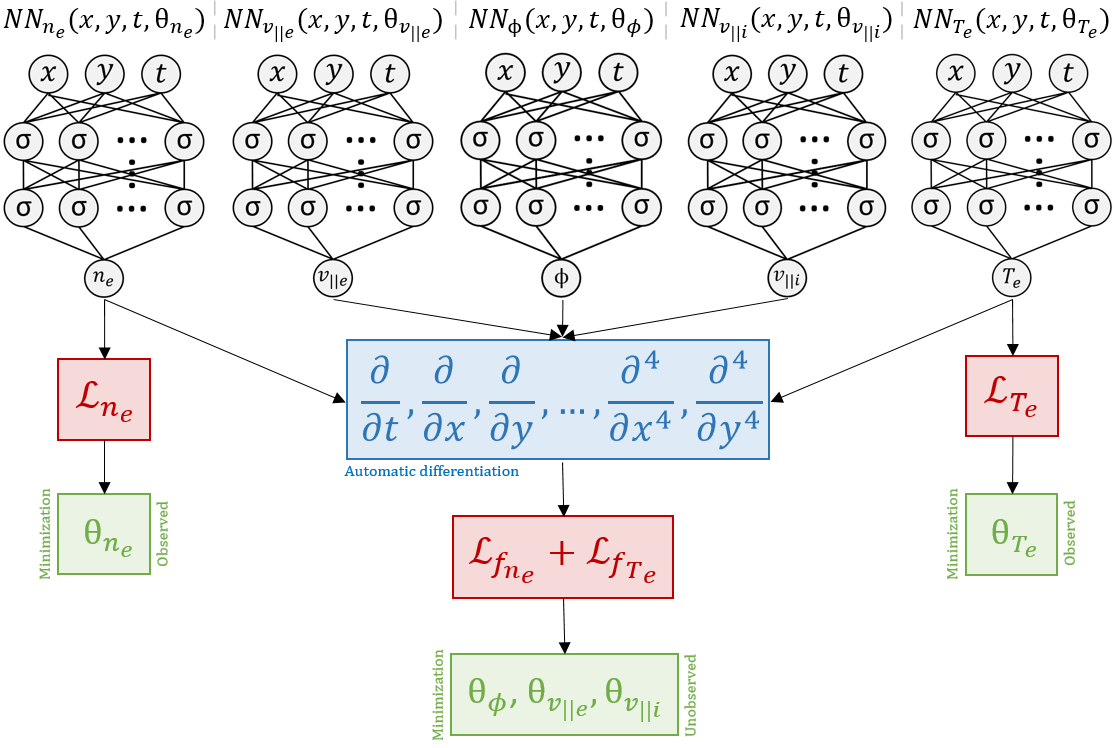}
\caption{\label{PlasmaPINN} Visualization of the physics-informed framework with individual networks being selectively trained against loss functions comprising both partial observations, $\mathcal{L}_{n_e}$ and $\mathcal{L}_{T_e}$, and reduced theory, $\mathcal{L}_{f_{n_e}}$ and $\mathcal{L}_{f_{T_e}}$, to infer unobserved turbulent dynamics. All spatial gradients and time derivatives in $f_{n_e}$ and $f_{T_e}$ are represented using automatic differentiation of each individual variable's network which in practice extends the size of the computation graph being evaluated. To handle reduced 2-dimensional data from the 3-dimensional synthetic plasma, the z-coordinate is removed from the networks for simplicity and as a test for determining the minimal information necessary to learn $\phi$. If noisy data is observed, then $\theta_{n_e}$ (and $\theta_{T_e}$ if $T_e$ measurements are available) should be additionally trained against $\mathcal{L}_{f_{n_e}}$ (and $\mathcal{L}_{f_{T_e}}$).}
\end{figure*}

In actuality, the repeated differentiation and summation of networks to construct every single term in the partial differential equations trained upon subsequently constructs a far larger resultant computation graph representing the collective model beyond the 5 hidden layers in each dynamical variable's individual network. The cumulative graph is therefore a truly deep approximation of the turbulent plasma theory. Physical constraints are learned by the networks via minimization of ascribed loss functions encompassing both limited measurements of the plasma and two-fluid turbulence model. In particular, partial observations of the simulated plasma consist of only $n_e$ and $T_e$ measurements of 2-dimensional spatial extent as visualized in Figure \ref{observed_dens_and_Te} over just 4 separate time slices (i.e. 2.9 $\mu$s). For reference, the synthetic plasma's fluctuations have an approximate autocorrelation time of 1.5 $\mu$s and radial autocorrelation length of 0.4 cm. The narrow temporal extent of the strongly fluctuating $n_e$ observations at a local spatial point is further visualized in Figure \ref{observed_1D_dens_time}. Properties of all other dynamical variables in the 6-field turbulence theory are taken to be unknown, and the networks are simultaneously optimized against the drift-reduced Braginskii equations and observed data to better approximate the unmeasured quantities. To be precise, partial synthetic measurements are learned by training the $n_e$ and $T_e$ networks against the average $\mathcal{L}_2$-norm of their respective relative errors
\begin{equation}\label{eq:L_n_DotGDBH}
    \mathcal{L}_{n_e} = \frac{1}{N_0}\sum_{i=1}^{N_0} \lvert n^*_{e}(x^i_0,y^i_0,z^i_0,t^i_0) - n^i_{e,0} \rvert^2,
\end{equation}
\begin{equation}\label{eq:L_Te_DotGDBH}
    \mathcal{L}_{T_e} = \frac{1}{N_0}\sum_{i=1}^{N_0} \lvert T^*_{e}(x^i_0,y^i_0,z^i_0,t^i_0) - T^i_{e,0} \rvert^2,
\end{equation}
where $\lbrace x_0^i,y_0^i,z_0^i,t_0^i,n_{e,0}^i,T_{e,0}^i\rbrace^{N_0}_{i=1}$ correspond to the set of observed data and the variables $n^*_e$ and $T^*_e$ symbolize predicted electron density and temperature, respectively, by the networks. The theory enforcing physical constraints in the deep learning framework is expressed by evaluating the individual terms in the model by differentiating the neural networks with respect to input spatiotemporal coordinates via application of chain rule through automatic differentiation \cite{tensorflow2015-whitepaper}. Correspondingly, model loss functions are embedded during training by recasting the evolution equations of \eqref{eq:nDotGDBH} and \eqref{eq:TeDotGDBH} in the following implicit form

\begin{eqnal}\label{eq:f_n_DotGDBH}
f_{n_e} &\coloneqq -\d{^e n}{t} -\frac{2c}{B}\left[n\curv{\phi}-\frac{1}{e}\curv{p_e}\right]\\ &\quad-n\delpar{\vpe} +\nSrcN+\mathcal{D}_{n},
\end{eqnal}

\begin{eqnal}\label{eq:f_Te_DotGDBH}
f_{T_e} &\coloneqq -\d{^e T_e}{t} + \frac{2T_e}{3n}\left[\d{^e n}{t} + \frac{1}{T_e}\delpar \kappa^e_\parallel \delpar T_e \right.\\& \left. + \frac{5n}{m_e \Omega_e} \curv{T_e} + \eta_\parallel \frac{\cur^2}{T_e} + \frac{0.71}{e}(\delpar{\cur} \right.\\ & \left. - \frac{\cur}{T_e}\delpar{T_e}) + \frac{1}{T_e} \enerSrceN \right] + \mathcal{D}_{T_e},
\end{eqnal}
and then further normalized into dimensionless form matching the numerical code as in \eqref{eq:normlnDotGDBH} and \eqref{eq:normlTeDotGDBH} \cite{francisquez2020fluid}. The normalized implicit formulation is vital to learning via optimization since all physical terms collectively sum to zero when the equations are ideally satisfied. These physical constraints provided by the unitless evolution equations of $n_e$ and $T_e$ from the two-fluid model are jointly optimized using loss functions defined by
\begin{equation}\label{eq:L_f_n_DotGDBH}
    \mathcal{L}_{f_{n_e}} = \frac{1}{N_f}\sum_{i=1}^{N_f} \lvert f^{*}_{n_e}(x^i_f,y^i_f,z^i_f,t^i_f) \rvert^2,
\end{equation}
\begin{equation}\label{eq:L_f_Te_DotGDBH}
    \mathcal{L}_{f_{T_e}} = \frac{1}{N_f}\sum_{i=1}^{N_f} \lvert f^{*}_{T_e}(x^i_f,y^i_f,z^i_f,t^i_f) \rvert^2,
\end{equation}
where $\lbrace x_f^i,y_f^i,z_f^i,t_f^i\rbrace^{N_f}_{i=1}$ denote the set of collocation points, and $f^{*}_{n_e}$ and $f^{*}_{T_e}$ are the null partial differential equations prescribed by \eqref{eq:f_n_DotGDBH} and \eqref{eq:f_Te_DotGDBH} in normalized form directly evaluated by the neural networks. Optimization against the applied theory is central to the methodology and enforces physical constraints in the deep learning framework by ensuring the networks respect the proposed plasma turbulence model. This enables fine tuning of the neural networks' weights and biases ($\theta$) by adjusting them in this generalized regression model to satisfy the physical laws governing the turbulent dynamics. The set of collocation points over which the partial differential equations are evaluated can be arbitrarily large and span any extent over the physical domain, but are taken in this example to correspond to the positions of the synthetic measurements being trained upon, i.e. $\lbrace x_0^i,y_0^i,z_0^i,t_0^i\rbrace^{N_0}_{i=1} = \lbrace x_f^i,y_f^i,z_f^i,t_f^i\rbrace^{N_f}_{i=1}$. 
It should be once again noted that the only observed dynamical quantities in these equations are 2-dimensional views of $n_e$ and $T_e$ without any explicit information about boundary conditions nor initializations. All analytic terms encoded in these equations including high-order operators are computed exactly by the neural networks without any approximation (e.g. linearization) nor discretization. This machine learning framework with a collocation grid of arbitrarily high resolution uses a continuous spatiotemporal domain without time-stepping nor finite difference schema in contrast with the numerical code. To handle 2-dimensional data, we  assume slow variation of dynamics in the $z$-coordinate and effectively set all parallel derivatives to zero (i.e. $\frac{\partial}{\partial z}\rightarrow 0$). Notwithstanding, parallel flows and Ohmic heating terms in the model are still kept. If measurements in the $z$-direction are available or more collocation points utilized during training with observational data of reduced dimensionality, this procedure may be relaxed---it is partly a trade-off between computational fidelity and stability. It is noteworthy that the temporal resolution of the data observed by the neural networks is several orders of magnitude lower than the time steps taken by the finite difference solver as required for numerical stability, i.e. $\Delta^* t \gg \Delta t$. Also, if sought, training on data sets viewed at oblique angles in 3-dimensional space over long macroscopic timescales can be easily performed via segmentation of the domain and parallelization, but a limited spatial view with reduced dimensionality is taken to emulate experimental conditions for nominally field-aligned observations \cite{Zweben_2017} and theoretically test what information is indispensable to learn unobserved turbulent dynamics. 

Loss functions are optimized with mini-batch sampling where $N_0 = N_f = 500$ using stochastic gradient descent via Adam \cite{kingma2014adam} and L-BFGS---a quasi-Newton optimization algorithm \cite{10.5555/3112655.3112866}--- for 20 hours over 32 cores on Intel Haswell-EP processors which corresponds to approximately 8694 full iterations over both optimizers. If observing noisy data, we find that expanding to larger sample sizes with $N_0 = N_f = 2500$ and training solely via L-BFGS is optimal for learning. Removing $\mathcal{L}_{f_{n_e}}$ and $\mathcal{L}_{f_{T_e}}$, i.e. setting $N_f = 0$, from the optimization process would correspond to training of classical neural networks without knowledge of the underlying governing equations and therefore an inability to learn unknown quantities. Setting $N_0 = 0$ instead while providing initial and boundary conditions for all dynamical variables would alternatively correspond to solving the equations directly via neural networks. Overall, priming networks by firstly training in stages on observed data or prior constraints is useful to enhance stability and convergence in this multi-objective task. Additionally, encoding domain expertise such as subsonic bounds on parallel flows or regularizing temperature to be strictly positive via suitable output activation functions can assist training by constraining the admissible landscape of solutions. Neural networks constructed in this way can intrinsically abide by physical laws which is especially useful to uncover unknowns like $\vpe$ and $T_i$.

A fundamental goal in computational plasma modelling is determining the minimum complexity necessary (and no less) to develop sufficiently predictive tokamak simulations. With sparse availability of measurements in fusion experiments, designing diagnostic techniques for uncovering such information is crucial. On this point, we emphasize that training is from scratch over just a single synthetic plasma discharge with no extraneous validation nor testing sets required since overfitting is technically not encountered in this physics-informed paradigm. The multi-network deep learning framework simply utilizes a single set of $n_e$ and $T_e$ measurements over a period of microseconds which corresponds to the small data regime of machine learning. Merging partial observational data of $n_e$ and $T_e$ along with physical laws in the form of partial differential equations governing the time-dependent evolution of $n_e$ and $T_e$ sufficiently constrains the set of admissible solutions for the previously unknown nonlinear mappings the neural networks ultimately learn. It is also quite general: due to quasineutrality, no significant adjustments are necessary to generalize the technique when multiple ions and impurities may be present in boundary plasmas beyond the inclusion of appropriate collisional drifts and sources in multi-species plasmas \cite{multi_species}. This deep learning technique for diagnosing turbulent fields is hence easily transferable which permits its systematic application across magnetic confinement fusion experiments whereby the underlying physical model fundamental to the turbulent transport is consistent. The framework sketched can also be readily extended to different settings in the interdisciplinary study (both numerical and experimental) of magnetized collisional plasmas in propulsion engines and astrophysical environments.

\section{\label{sec:level4} Numerical Experiments}

\begin{figure}[ht]
\includegraphics[width=1.0\linewidth]{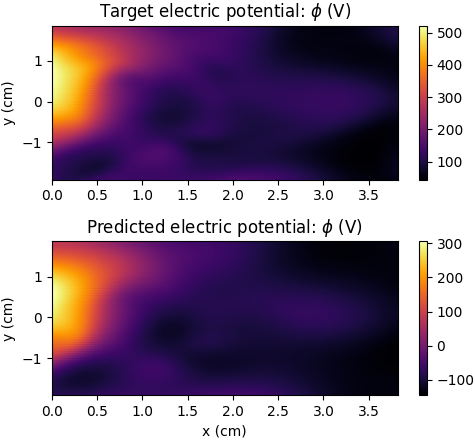}
\caption{\label{learned_phi} The synthetic plasma's unobserved electric potential (top) is learned approximately up to an additive constant as predicted by the physics-informed neural network (bottom).}
\end{figure}

\begin{figure}[ht]
\includegraphics[width=1.00\linewidth]{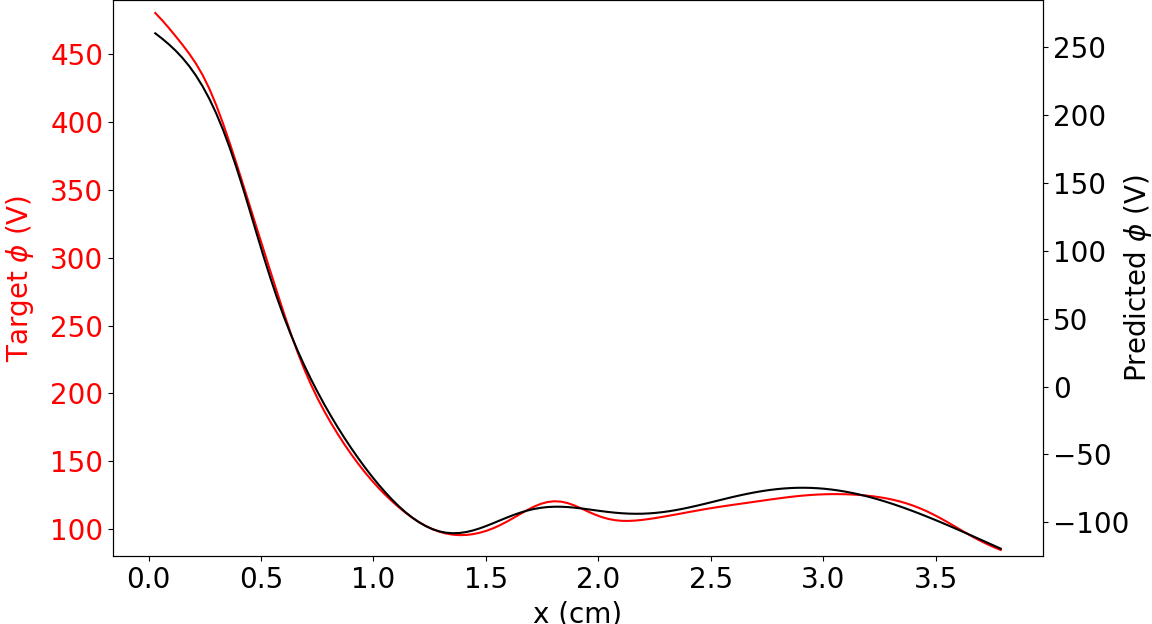}
\caption{\label{1d_phi} 1-dimensional radial profile of the true and predicted $\phi$ at $[y = 0.0 \text{ cm}, z = -28.1 \text{ cm}, t = 229.9 \ \mu\text{s}]$ corresponds to a slice of Figure \ref{learned_phi}. The ordinates have identical ranges spanning exactly 410 V with equivalent spacing for direct comparison.}
\end{figure}

Accurate turbulent edge electric field fluctuation characterization is particularly significant to magnetic confinement fusion devices. By constraining the deep learning framework with the two-fluid turbulence theory and strikingly little empirical information in the form of partial 2-dimensional observations of $n_e$ and $T_e$, we find that physics-informed neural networks can accurately learn the plasma's turbulent electric potential without ever having observed it, as displayed in Figures \ref{learned_phi} and \ref{1d_phi}. 

\begin{figure}[ht]
\includegraphics[width=1.0\linewidth]{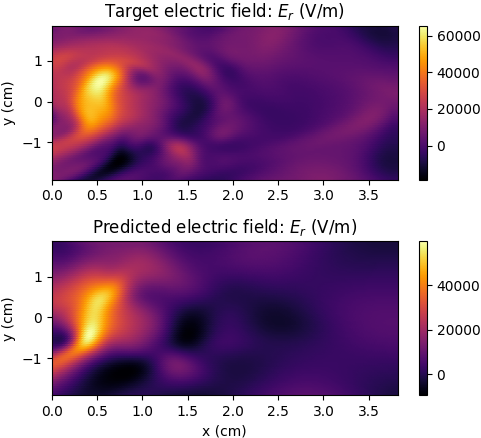}
\caption{\label{learned_E_r} The learned turbulent $E_r$ (bottom) closely agrees with the magnitude and structure of the true $E_r$ (top) despite $\gvort$, $\vpe$, $\vpi$, and $T_i$ being initially unknown.}
\end{figure}

It is notable that despite there being no knowledge of $\gvort, \vpe, \vpi,$ nor $T_i$ (i.e. multiple unknowns existing in the partial differential equations and \eqref{BVP_gvort_phi} never even being directly invoked), the electric field is nonetheless learned consistently with the physical theory encoded  by the plasma turbulence model. Since $\phi$ is a gauge-invariant quantity exact up to an additive constant, it is accordingly uncovered up to a scalar offset which varies randomly due to the stochastic nature of the optimization employed in the machine learning framework. This difference physically arises because no direct boundary information was enforced upon the neural network when learning $\phi$, although it could be straightforwardly implemented. By contrast, the numerical code imposed zero potential on the outer walls of the simulation domain. General agreement in both magnitude and structure in the learned radial electric field is evident in Figure \ref{learned_E_r} with an average absolute error of 26.19 V/cm while $\mathcal{L}_{f_{n_e}}$ and $\mathcal{L}_{f_{T_e}}$ are $3.101 \times 10^{-1}$ and $3.743 \times 10^{-1}$, respectively, after 8694 full iterations. In physical units, $\mathcal{L}_{f_{n_e}}$ and $\mathcal{L}_{f_{T_e}}$ would be $3.472 \times 10^{17} \text{ cm}^{-3}\text{/s}$ and $1.273 \times 10^{6}  \text{ eV}\text{/s}$.

\begin{figure}
\includegraphics[width=1.0\linewidth]{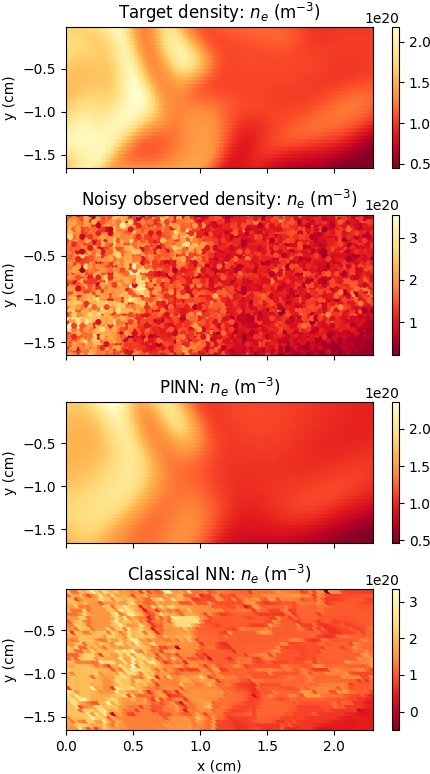}
\caption{\label{noisy_dens} The physics-informed deep learning framework is capable of recovering the true $n_e$ despite strong Gaussian noise ($\sigma = 0.25$) present. The classical solution corresponds to a standard feed-forward neural network where $N_f = 0$.}
\end{figure}

To emphasize the experimental practicality of these results and a surprising feature discovered, we note that the radial and poloidal electric fields can actually be partly learned and reproduced with just partial measurements of $n_e$ without any observations of $T_e$. Therefore, while $T_e$ observations provide stringent constraints on the solution space for improved convergence, they are not strictly necessary---knowing $T_e$ primarily stabilizes optimization while learning $\phi$. On this point, we note that if only observing density fluctuations with strong Gaussian noise, one could still largely recover the unmeasured electric field and even resolve the partial observations. Namely, given just highly noisy measurements as in Figure \ref{noisy_dens}, one can learn the true turbulent density in this physics-informed deep learning framework to subsequently infer the unmeasured electric field. If $E_r$ was already known, this technique could then precisely check the validity of the reduced turbulence theory against observations from experiment or kinetic simulations \cite{Mathews2021_FluidvGk}. But, if using a standard feed-forward neural network, one must be careful with convergence since the objective of simply minimizing $\mathcal{L}_{n_e}$ without sufficient regularization, as innately provided by $\mathcal{L}_{f_{n_e}}$, can result in overfitting of noisy data.

\begin{figure}
\includegraphics[width=1.0\linewidth]{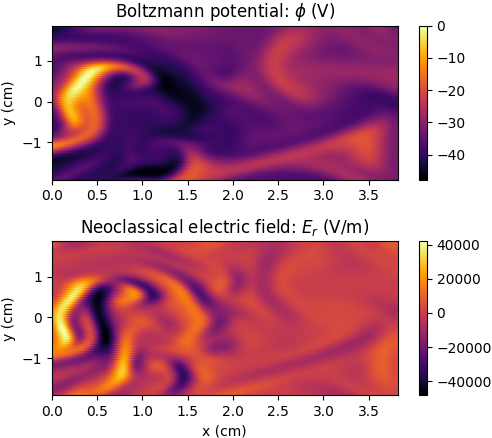}
\caption{\label{analytic_E_r} Estimates of the turbulent $\phi$ and $E_r$ as expected by the Boltzmann model, $n_e(\phi) = n_e(\phi_0) e^{e(\phi - \phi_0)/T_e}$ \cite{Callen_adiabatic}, or neoclassical estimates, $\nabla \phi = -\nabla p_i/Zn_i e$ \cite{Viezzer_2013}, where $Z=1$ for deuterium ions, yield markedly errant predictions in both structure and magnitude when compared to the true values displayed at the top of Figures \ref{learned_phi} and \ref{learned_E_r}.}
\end{figure}

For comparison, classical and oft-employed models for calculating the electric potential with adiabatic electrons such as the Boltzmann relation \cite{Callen_adiabatic} fail when computing perpendicular turbulent field fluctuations. Alternative approximations of $E_r$ from simple ion pressure balance as expected neoclassically \cite{Viezzer_2013} would yield highly incorrect estimates of the turbulent electric field, too. Such methods ordinarily used in magnetic confinement fusion are only applicable to discerning equilibrium fields and dynamics parallel to the magnetic field in steady-state scenarios, but are erroneous in the analysis of microturbulence in nonquiescent plasmas as markedly observed when comparing Figure \ref{analytic_E_r} to the true $\phi$ and $E_r$ as plotted in Figures \ref{learned_phi} and \ref{learned_E_r}, respectively. Our deep learning technique based upon drift-reduced Braginskii theory therefore provides a novel way to accurately measure the turbulent electric field in edge plasmas from just the electron pressure. As a further point of contrast compared to classical techniques, it is important to note that the inverse learning scenario is not possible. In particular, given observations of $\phi$ and $T_e$, one cannot simply infer the turbulent $n_e$ fluctuations with the machine learning framework outlined. This one-way nature in learning indicates a division exists between the two pathways when attempting to constrain the admissible solutions of \eqref{eq:L_f_n_DotGDBH} and \eqref{eq:L_f_Te_DotGDBH} to uncover unknown nonequilibrium dynamics. Training is thus unidirectional and representative of asymmetries extant in the partial data and turbulence theory being learnt via optimization in the physics-informed model. 

\begin{table}[ht]
{\centering
\renewcommand{\arraystretch}{1.0}
\includegraphics[width=1.0\linewidth]{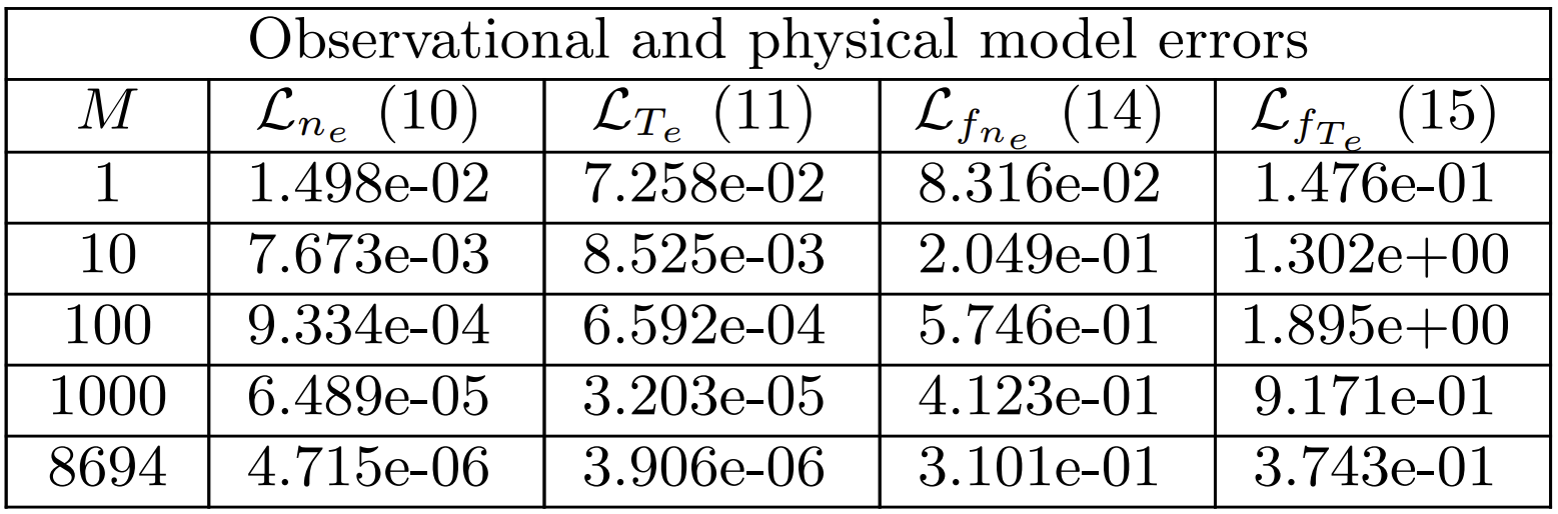}
\caption{\label{loss_tabulate} Each normalized loss function optimized in the machine learning framework is tabulated after $M$ full iterations, where $M = 8694$ corresponds to the final iteration after 20 hours of training against both the partial observations of $n_e$ and $T_e$ and their implicit evolution equations.}
}
\end{table}

To better interpret the learning process, we tabulate the normalized loss functions being trained upon after $M$ full iterations by the optimizers in Table \ref{loss_tabulate}. After 1 iteration, \eqref{eq:L_f_n_DotGDBH} and \eqref{eq:L_f_Te_DotGDBH} are relatively small in magnitude, and this would correspond to a trivial result satisfying the partial differential equations given the nonuniqueness of its solutions. As training progresses, observational data is better captured in the deep learning framework and the neural networks proceed to converge toward the sought solution as opposed to trivial ones. A difference in the rates of learning for $n_e$, $T_e$, and $\phi$ also exist since the electric field is learned implicitly via the model instead of being trained upon directly. Each individual loss function being optimized therefore does not necessarily decrease perfectly monotonically, but it is instead the collective training against partial differential equations in conjunction with observational data that is key. Namely, while there are many potential solutions to \eqref{eq:f_n_DotGDBH} and \eqref{eq:f_Te_DotGDBH}---and while they may be more easily satisfied by trivial solutions---the observational data compels the optimizer towards the physical solution of the observed plasma. In scenarios where inconsistencies in the true and learned model $E_r$ exist, one can repurpose this machine learning framework to iteratively test and thereby discover the {\it correct} partial differential equations altogether by quantitatively examining the proposed model's consistency with observations as in Table \ref{loss_tabulate}. For example, the analytic form of reduced source models in fluid theories \cite{Thrysoe_2018,Thrysoe_2020} can be inserted in the physics-informed deep learning framework to account for local turbulent ionization and inelastic collisions with kinetic neutrals by observing such measurements of $n_e$, $T_e$, and $\phi$ in global simulations \cite{Wersal_2015} and experiments \cite{Mathews2021}. 

\section{\label{sec:level5} Conclusion}

These results illustrate a novel physics-informed deep learning paradigm with the capacity to learn unknown dynamics in a multi-field turbulent transport model broadly relevant to magnetized collisional plasmas. We specifically demonstrate the ability for neural networks to determine unobserved turbulent electric fields consistent with the drift-reduced Braginskii equations from partial electron pressure observations which is not otherwise possible using standard analytic techniques. This can be applied to infer field fluctuations that may be difficult to measure or when sought plasma diagnostics are simply lacking. On the other hand, if experimental electric field measurements exist, then the quantitative validity of the plasma turbulence model embedded in the neural networks can be expressly assessed. This technique is also quite robust since, due to quasineutrality, it can be used to study ionized gases in magnetized environments with multiple ions and impurities present as commonly found in astrophysical settings and fusion energy and space propulsion systems. From a mathematical physics standpoint, it is significant that nonlinear dynamics can be accurately recovered from partial data and theory in a 6-field model. Inferring completely unknown turbulent fields from just 2-dimensional measurements and representations of the evolution equations given by \eqref{eq:f_n_DotGDBH} and \eqref{eq:f_Te_DotGDBH} demonstrates a massive reduction in the original 3-dimensional turbulence model indicating redundancy and the existence of reduced theory characterizations. Going forward, this framework has the capability to be generalized (e.g. to learn $T_e$, $T_i$, $\vpe$, and $\vpi$ in addition to $\phi$ using just 1-dimensional $n_e$ measurements) and transform how turbulence theories are systematically and quantitatively validated in both plasma simulations and experiments. The interpretable physics-informed methodology outlined is also transferable across models (e.g. collisionless fluids, gyrokinetic, electromagnetic, atomic physics) and complex geometries. Furthermore, known limitations and {\it unknown} corrections to Braginskii's theory exist \cite{Catto-paper} which can be introduced in the deep learning framework to automate efficient testing and discovery of reduced plasma turbulence models when high fidelity data is observed. These extensions in theory and computing will be focused upon in future works.

 

\appendix
\section{Normalization of plasma theory}

Converting the drift-reduced Braginskii equations from physical units to a normalized form is essential to numerically solve the model in both finite difference solvers and physics-informed machine learning codes. For completeness, the full normalization procedure is carried out whereby the physical variables and all associated quantities are transformed according to \cite{francisquez2020fluid}
\begin{align} \label{eq:normSMT1}
\begin{split}
  n &\leftarrow n/\nRef, \\
  \phi &\leftarrow \phi/\phi_0, \\
\end{split}
\quad
\begin{split}
  T_{s} &\leftarrow T_{s}/T_{s0}, \\
  v_{\parallel s} &\leftarrow v_{\parallel s}/c_{s0}, \\
\end{split}
\end{align}
where $n_0 = 5 \times 10^{19} \text{ m}^{-3}$, $T_{s0} = 25 \text{ eV}$, $c_{s0}^2 = T_{s0}/m_i$, $\phi_0 = \BRef a^2/c\tRef$, and $\tRef = \sqrt{a_0 R_c/2}/c_{e0}$ is the interchange-like reference timescale. To match experimental edge parameters of the Alcator C-Mod tokamak, $\BRef = B_{axis} R_0 /(R_0 + a_0)$ and $R_c = R_0 + a_0$. This in turn defines the following dimensionless constants

\begin{eqnal}
\!\begin{aligned}
  \epa &= \frac{2a}{R_c}, \\
  \epv &= \frac{c_{e0}\tRef}{R_c}, \\
  \tau &= \frac{\TiRef}{\TeRef},
\end{aligned}
\qquad
\!\begin{aligned}
  \kappai &= 3.9\frac{2}{3}\frac{\TiRef\tRef}{m_i R_c^2\nu_{i0}}, \\
  \eta &= 0.51\nu_{e0}\tRef, \\
  \kappae &= 3.2\frac{2}{3}\frac{\TeRef\tRef}{\nu_{e0}m_e R_c^2},
\end{aligned}
\qquad
\!\begin{aligned}
  \alphad &= \frac{\TeRef c\tRef}{e\BRef a^2}, \\
  \epg &= \frac{0.08\tau}{\nu_{i0} t_0}, \\
  \epge &= \frac{0.73}{12\nu_{e0}t_0}
\end{aligned}
\end{eqnal}
where $c$ and $\nu_{s 0}$ denote the speed of light and collision rate \cite{Huba2013}, respectively. The spatiotemporal coordinates are normalized by the following conversions
\begin{align} \label{eq:normXYZT}
\begin{split}
  x &\leftarrow x/a_0, \\
  z &\leftarrow z/R_0, \\
\end{split}
\quad
\begin{split}
  y &\leftarrow y/a_0, \\
  t &\leftarrow t/t_0. \\
\end{split}
\end{align}
Under these set of transformations, the normalized two-fluid equations numerically solved are
\begin{eqnal}\label{eq:normlnDotGDBH}
\d{^e\ln n}{t} &= -\epa\left[\curv{\phi}-\alphad\frac{\curv{p_e}}{n}\right]\\ &\quad -\epv\delpar{\vpe}+\frac{1}{n}\nSrcN+\mathcal{D}_{\ln n}
\end{eqnal}
\begin{eqnal}\label{eq:normwDotGDBH}
\pd{\gvort}{t} &= \curv{p_e}+\tau\curv{p_i}+\frac{\epv}{\alphad\epa}\delpar{\cur}-\epg\curv{G_i} \\
&\quad-\div{\lbrace\frac{n}{B^3}\left[\phi,\gradperp{\phi}+\tau\alphad\frac{\gradperp{p_i}}{n}\right]+ \\
&\quad\sqrt{\tau}\epv\frac{n}{B^2}\vpi\delpar{\left(\gradperp{\phi}+\tau\alphad\frac{\gradperp{ p_i}}{n}\right)}\rbrace}+\mathcal{D}_{\gvort}
\end{eqnal}
\begin{eqnal}\label{eq:normvpeDotGDBH}
\d{^e\vpe}{t} &= \frac{m_i}{m_e}\epv\left(\frac{1}{\alphad}\delpar{\phi}-\frac{\delpar{p_e}}{n}-0.71\delpar{T_e}\right) \\ &\quad +4\epv\epge\frac{m_i}{m_e}\frac{\delpar{G_e}}{n} +\epa\alphad T_e\curv{\vpe} \\ &\quad +\eta\frac{\cur}{T_e^{3/2}} +\momSrce +\mathcal{D}_{\vpi}
\end{eqnal}
\begin{eqnal}\label{eq:normvpiDotGDBH}
\d{^i\vpi}{t} &= -\frac{\epv}{\sqrt{\tau}}\left(\frac{1}{\alphad}\delpar{\phi}+\tau\frac{\delpar{p_i}}{n}-0.71\delpar{T_e}\right) \\
&\quad +\frac{4\epv\epg}{\sqrt{\tau}}\frac{\delpar{G_i}}{n} -\epa\tau\alphad T_i\curv{\vpi} \\ &\quad -\frac{m_e}{m_i}\frac{\eta}{\sqrt{\tau}}\frac{\cur}{T_e^{3/2}} +\momSrci+\mathcal{D}_{\vpi}
\end{eqnal}
\begin{eqnal}\label{eq:normlTeDotGDBH}
\d{^e\ln T_e}{t} = \frac{5}{3}\epa\alphad\curv{T_e}+\frac{\kappa^e}{p_e}\delpar{T_e^{7/2}\delpar{\ln T_e}} \\  +\frac{2}{3}\left\lbrace -\epa\left[\curv{\phi}-\alphad\frac{\curv{p_e}}{n}\right]-\epv\delpar{\vpe} \right.\\
\left.+\frac{1}{n}\left[0.71\epv\left(\delpar{\cur}-\cur\delpar{\ln T_e}\right) \right.\right.\\
\left.\left. +\frac{m_e}{m_i}\eta\frac{\cur^2}{T_e^{5/2}}\right]\right\rbrace +\frac{2}{3}\frac{1}{p_e}\enerSrceN +\mathcal{D}_{\ln T_e}
\end{eqnal}
\begin{eqnal}\label{eq:normlTiDotGDBH}
\d{^i\ln T_i}{t} = -\frac{5}{3}\tau\epa\alphad \curv{T_i}+\frac{\kappa^i}{p_i}\delpar{T_i^{7/2}\delpar{\ln T_i}} \\ +\frac{2}{3}\left\lbrace -\epa\left[\curv{\phi}-\alphad\frac{\curv{p_e}}{n}\right] -\sqrt{\tau}\epv\delpar{\vpi} \right.\\
\left. +\epv\frac{\delpar{\cur}}{n}\right\rbrace  +\frac{2}{3}\frac{1}{p_i}\enerSrciN+\mathcal{D}_{\ln T_i},
\end{eqnal}
where the normalized diffusivities applied for all dynamical variables are $\chi_x = -4.54 \times 10^{-10}$, $\chi_y = -1.89 \times 10^{-9}$, and $\chi_z = -8.91 \times 10^{-3}$. The normalized evolution equations given by \eqref{eq:normlnDotGDBH} and \eqref{eq:normlTeDotGDBH} are the physical model constraints learnt in the machine learning framework as detailed above. A few subtle yet importance differences exist between the physical theory posed and the construction of the synthetic plasma. One deviation between the theorized plasma and the one produced computationally is that the numerical code actually evolves the logarithmic form of $n$, $T_e$, and $T_i$ to enforce positivity and the high order diffusion operators act on these logarithmic quantities, too. While equivalent analytically, this choice numerically forces the drift-reduced Braginskii equations to be posed and solved in non-conservative form by the finite difference solver. Consequent errors due to numerical approximation can manifest as unexpected artificial sources or sinks in the simulation domain \cite{francisquez2020fluid}. In addition, simulation boundaries applied in practice only approximately satisfy the zero flux conditions when employing even- and odd-symmetry conditions on a cell-centered discretized grid \cite{francisquez2020fluid}. These computational discrepancies can cause potential misalignment between inferred dynamics using idealized theory and numerical modelling of the synthetic plasma's turbulent fields. Physics-informed deep learning can overcome these numerical limitations when representing plasma theory since positivity can be intrinsically encoded in the network plus it employs a continuous spatiotemporal domain and the nonlinear continuum equations represented by \eqref{eq:f_n_DotGDBH} and \eqref{eq:f_Te_DotGDBH} are consequently evaluated exactly up to computer precision \cite{Raissi_JMLR}. Unphysical numerical dissipation in observational data can therefore present deviations from reflecting the sought plasma theory, but reasonable agreement is nevertheless found when analyzing the synthetic measurements with the partial differential equations embedded in the physics-informed machine learning framework.
 
 

\begin{acknowledgments}
\noindent We wish to thank P. Perdikaris, S. Wang, and A.Q. Kuang for insights shared and helpful discussions. All simulations presented and codes run are performed using MIT's Engaging cluster and we are grateful for the team's maintenance and assistance with computing resources. The work is supported by the Natural Sciences and Engineering Research Council of Canada (NSERC) by the doctoral postgraduate scholarship (PGS D), Manson Benedict Fellowship, and the U.S. Department of Energy (DOE) Office of Science under the Fusion Energy Sciences program by contracts DE-SC0014264, DE-FC02-08ER54966, DE-FG02-04ER54742, and DE-AC52-07NA27344. Relevant data input files and codes can be found on Github (\href{github.com/AbhilashMathews/PlasmaPINNs}{\color{blue}github.com/AbhilashMathews/PlasmaPINNs}) and the repository will become public upon publication.
\end{acknowledgments}

\nocite{*}
\bibliographystyle{apsrev4-2}
\bibliography{main.bib}

\begin{thebibliography}{55}%
\makeatletter
\providecommand \@ifxundefined [1]{%
 \@ifx{#1\undefined}
}%
\providecommand \@ifnum [1]{%
 \ifnum #1\expandafter \@firstoftwo
 \else \expandafter \@secondoftwo
 \fi
}%
\providecommand \@ifx [1]{%
 \ifx #1\expandafter \@firstoftwo
 \else \expandafter \@secondoftwo
 \fi
}%
\providecommand \natexlab [1]{#1}%
\providecommand \enquote  [1]{``#1''}%
\providecommand \bibnamefont  [1]{#1}%
\providecommand \bibfnamefont [1]{#1}%
\providecommand \citenamefont [1]{#1}%
\providecommand \href@noop [0]{\@secondoftwo}%
\providecommand \href [0]{\begingroup \@sanitize@url \@href}%
\providecommand \@href[1]{\@@startlink{#1}\@@href}%
\providecommand \@@href[1]{\endgroup#1\@@endlink}%
\providecommand \@sanitize@url [0]{\catcode `\\12\catcode `\$12\catcode
  `\&12\catcode `\#12\catcode `\^12\catcode `\_12\catcode `\%12\relax}%
\providecommand \@@startlink[1]{}%
\providecommand \@@endlink[0]{}%
\providecommand \url  [0]{\begingroup\@sanitize@url \@url }%
\providecommand \@url [1]{\endgroup\@href {#1}{\urlprefix }}%
\providecommand \urlprefix  [0]{URL }%
\providecommand \Eprint [0]{\href }%
\providecommand \doibase [0]{https://doi.org/}%
\providecommand \selectlanguage [0]{\@gobble}%
\providecommand \bibinfo  [0]{\@secondoftwo}%
\providecommand \bibfield  [0]{\@secondoftwo}%
\providecommand \translation [1]{[#1]}%
\providecommand \BibitemOpen [0]{}%
\providecommand \bibitemStop [0]{}%
\providecommand \bibitemNoStop [0]{.\EOS\space}%
\providecommand \EOS [0]{\spacefactor3000\relax}%
\providecommand \BibitemShut  [1]{\csname bibitem#1\endcsname}%
\let\auto@bib@innerbib\@empty
\bibitem [{\citenamefont {{Carreras}}(1997)}]{Carreras_1997}%
  \BibitemOpen
  \bibfield  {author} {\bibinfo {author} {\bibfnamefont {B.~A.}\ \bibnamefont
  {{Carreras}}},\ }\href@noop {} {\bibfield  {journal} {\bibinfo  {journal}
  {IEEE Transactions on Plasma Science}\ }\textbf {\bibinfo {volume} {25}},\
  \bibinfo {pages} {1281} (\bibinfo {year} {1997})}\BibitemShut {NoStop}%
\bibitem [{\citenamefont {Chapman}\ and\ \citenamefont
  {Morris}(2019)}]{Chapman2019UKAEACT}%
  \BibitemOpen
  \bibfield  {author} {\bibinfo {author} {\bibfnamefont {I.~T.}\ \bibnamefont
  {Chapman}}\ and\ \bibinfo {author} {\bibfnamefont {A.~W.}\ \bibnamefont
  {Morris}},\ }\href@noop {} {\bibfield  {journal} {\bibinfo  {journal}
  {Philosophical transactions. Series A, Mathematical, physical, and
  engineering sciences}\ }\textbf {\bibinfo {volume} {377}} (\bibinfo {year}
  {2019})}\BibitemShut {NoStop}%
\bibitem [{\citenamefont {Federici}\ \emph {et~al.}(2001)\citenamefont
  {Federici}, \citenamefont {Skinner}, \citenamefont {Brooks}, \citenamefont
  {Coad}, \citenamefont {Grisolia}, \citenamefont {Haasz}, \citenamefont
  {Hassanein}, \citenamefont {Philipps}, \citenamefont {Pitcher}, \citenamefont
  {Roth}, \citenamefont {Wampler},\ and\ \citenamefont
  {Whyte}}]{Federici_2001}%
  \BibitemOpen
  \bibfield  {author} {\bibinfo {author} {\bibfnamefont {G.}~\bibnamefont
  {Federici}}, \bibinfo {author} {\bibfnamefont {C.}~\bibnamefont {Skinner}},
  \bibinfo {author} {\bibfnamefont {J.}~\bibnamefont {Brooks}}, \bibinfo
  {author} {\bibfnamefont {J.}~\bibnamefont {Coad}}, \bibinfo {author}
  {\bibfnamefont {C.}~\bibnamefont {Grisolia}}, \bibinfo {author}
  {\bibfnamefont {A.}~\bibnamefont {Haasz}}, \bibinfo {author} {\bibfnamefont
  {A.}~\bibnamefont {Hassanein}}, \bibinfo {author} {\bibfnamefont
  {V.}~\bibnamefont {Philipps}}, \bibinfo {author} {\bibfnamefont
  {C.}~\bibnamefont {Pitcher}}, \bibinfo {author} {\bibfnamefont
  {J.}~\bibnamefont {Roth}}, \bibinfo {author} {\bibfnamefont {W.}~\bibnamefont
  {Wampler}},\ and\ \bibinfo {author} {\bibfnamefont {D.}~\bibnamefont
  {Whyte}},\ }\href {https://doi.org/10.1088/0029-5515/41/12/218} {\bibfield
  {journal} {\bibinfo  {journal} {Nuclear Fusion}\ }\textbf {\bibinfo {volume}
  {41}},\ \bibinfo {pages} {1967} (\bibinfo {year} {2001})}\BibitemShut
  {NoStop}%
\bibitem [{\citenamefont {Li}\ \emph {et~al.}(2020)\citenamefont {Li},
  \citenamefont {Xu}, \citenamefont {Hughes}, \citenamefont {Terry},
  \citenamefont {Sun},\ and\ \citenamefont {Wang}}]{Li2020}%
  \BibitemOpen
  \bibfield  {author} {\bibinfo {author} {\bibfnamefont {N.~M.}\ \bibnamefont
  {Li}}, \bibinfo {author} {\bibfnamefont {X.~Q.}\ \bibnamefont {Xu}}, \bibinfo
  {author} {\bibfnamefont {J.~W.}\ \bibnamefont {Hughes}}, \bibinfo {author}
  {\bibfnamefont {J.~L.}\ \bibnamefont {Terry}}, \bibinfo {author}
  {\bibfnamefont {J.~Z.}\ \bibnamefont {Sun}},\ and\ \bibinfo {author}
  {\bibfnamefont {D.~Z.}\ \bibnamefont {Wang}},\ }\href
  {https://doi.org/10.1063/1.5126884} {\bibfield  {journal} {\bibinfo
  {journal} {AIP Advances}\ }\textbf {\bibinfo {volume} {10}},\ \bibinfo
  {pages} {015222} (\bibinfo {year} {2020})}\BibitemShut {NoStop}%
\bibitem [{\citenamefont {Chang}\ \emph {et~al.}(2021)\citenamefont {Chang},
  \citenamefont {Ku}, \citenamefont {Hager}, \citenamefont {Churchill},
  \citenamefont {Hughes}, \citenamefont {K{\"o}chl}, \citenamefont {Loarte},
  \citenamefont {Parail},\ and\ \citenamefont {Pitts}}]{Chang2021}%
  \BibitemOpen
  \bibfield  {author} {\bibinfo {author} {\bibfnamefont {C.~S.}\ \bibnamefont
  {Chang}}, \bibinfo {author} {\bibfnamefont {S.}~\bibnamefont {Ku}}, \bibinfo
  {author} {\bibfnamefont {R.}~\bibnamefont {Hager}}, \bibinfo {author}
  {\bibfnamefont {R.~M.}\ \bibnamefont {Churchill}}, \bibinfo {author}
  {\bibfnamefont {J.}~\bibnamefont {Hughes}}, \bibinfo {author} {\bibfnamefont
  {F.}~\bibnamefont {K{\"o}chl}}, \bibinfo {author} {\bibfnamefont
  {A.}~\bibnamefont {Loarte}}, \bibinfo {author} {\bibfnamefont
  {V.}~\bibnamefont {Parail}},\ and\ \bibinfo {author} {\bibfnamefont {R.~A.}\
  \bibnamefont {Pitts}},\ }\href {https://doi.org/10.1063/5.0027637} {\bibfield
   {journal} {\bibinfo  {journal} {Physics of Plasmas}\ }\textbf {\bibinfo
  {volume} {28}},\ \bibinfo {pages} {022501} (\bibinfo {year}
  {2021})}\BibitemShut {NoStop}%
\bibitem [{\citenamefont {{Braginskii}}(1965)}]{Braginskii1965RvPP}%
  \BibitemOpen
  \bibfield  {author} {\bibinfo {author} {\bibfnamefont {S.~I.}\ \bibnamefont
  {{Braginskii}}},\ }\href@noop {} {\bibfield  {journal} {\bibinfo  {journal}
  {Reviews of Plasma Physics}\ }\textbf {\bibinfo {volume} {1}},\ \bibinfo
  {pages} {205} (\bibinfo {year} {1965})}\BibitemShut {NoStop}%
\bibitem [{\citenamefont {Kim}\ \emph {et~al.}(1993)\citenamefont {Kim},
  \citenamefont {Horton},\ and\ \citenamefont {Hamaguchi}}]{Kim-1993}%
  \BibitemOpen
  \bibfield  {author} {\bibinfo {author} {\bibfnamefont {C.-B.}\ \bibnamefont
  {Kim}}, \bibinfo {author} {\bibfnamefont {W.}~\bibnamefont {Horton}},\ and\
  \bibinfo {author} {\bibfnamefont {S.}~\bibnamefont {Hamaguchi}},\ }\href
  {https://doi.org/10.1063/1.860891} {\bibfield  {journal} {\bibinfo  {journal}
  {Physics of Fluids B: Plasma Physics}\ }\textbf {\bibinfo {volume} {5}},\
  \bibinfo {pages} {1516} (\bibinfo {year} {1993})}\BibitemShut {NoStop}%
\bibitem [{\citenamefont {Francisquez}(2018)}]{francisquez_thesis}%
  \BibitemOpen
  \bibfield  {author} {\bibinfo {author} {\bibfnamefont {M.}~\bibnamefont
  {Francisquez}},\ }\emph {\bibinfo {title} {Global Braginskii modeling of
  magnetically confined boundary plasmas}},\ \href@noop {} {Ph.D. thesis},\
  \bibinfo  {school} {Dartmouth College} (\bibinfo {year} {2018})\BibitemShut
  {NoStop}%
\bibitem [{\citenamefont {Francisquez}\ \emph {et~al.}(2017)\citenamefont
  {Francisquez}, \citenamefont {Zhu},\ and\ \citenamefont
  {Rogers}}]{Francisquez_2017}%
  \BibitemOpen
  \bibfield  {author} {\bibinfo {author} {\bibfnamefont {M.}~\bibnamefont
  {Francisquez}}, \bibinfo {author} {\bibfnamefont {B.}~\bibnamefont {Zhu}},\
  and\ \bibinfo {author} {\bibfnamefont {B.}~\bibnamefont {Rogers}},\ }\href
  {https://doi.org/10.1088/1741-4326/aa7f23} {\bibfield  {journal} {\bibinfo
  {journal} {Nuclear Fusion}\ }\textbf {\bibinfo {volume} {57}},\ \bibinfo
  {pages} {116049} (\bibinfo {year} {2017})}\BibitemShut {NoStop}%
\bibitem [{\citenamefont {Ricci}\ \emph {et~al.}(2012)\citenamefont {Ricci},
  \citenamefont {Halpern}, \citenamefont {Jolliet}, \citenamefont {Loizu},
  \citenamefont {Mosetto}, \citenamefont {Fasoli}, \citenamefont {Furno},\ and\
  \citenamefont {Theiler}}]{Ricci_2012}%
  \BibitemOpen
  \bibfield  {author} {\bibinfo {author} {\bibfnamefont {P.}~\bibnamefont
  {Ricci}}, \bibinfo {author} {\bibfnamefont {F.~D.}\ \bibnamefont {Halpern}},
  \bibinfo {author} {\bibfnamefont {S.}~\bibnamefont {Jolliet}}, \bibinfo
  {author} {\bibfnamefont {J.}~\bibnamefont {Loizu}}, \bibinfo {author}
  {\bibfnamefont {A.}~\bibnamefont {Mosetto}}, \bibinfo {author} {\bibfnamefont
  {A.}~\bibnamefont {Fasoli}}, \bibinfo {author} {\bibfnamefont
  {I.}~\bibnamefont {Furno}},\ and\ \bibinfo {author} {\bibfnamefont
  {C.}~\bibnamefont {Theiler}},\ }\href
  {https://doi.org/10.1088/0741-3335/54/12/124047} {\bibfield  {journal}
  {\bibinfo  {journal} {Plasma Physics and Controlled Fusion}\ }\textbf
  {\bibinfo {volume} {54}},\ \bibinfo {pages} {124047} (\bibinfo {year}
  {2012})}\BibitemShut {NoStop}%
\bibitem [{\citenamefont {Thrys\o{}e}\ \emph {et~al.}(2018)\citenamefont
  {Thrys\o{}e}, \citenamefont {L\o{}iten}, \citenamefont {Madsen},
  \citenamefont {Naulin}, \citenamefont {Nielsen},\ and\ \citenamefont
  {Rasmussen}}]{Thrysoe_2018}%
  \BibitemOpen
  \bibfield  {author} {\bibinfo {author} {\bibfnamefont {A.~S.}\ \bibnamefont
  {Thrys\o{}e}}, \bibinfo {author} {\bibfnamefont {M.}~\bibnamefont
  {L\o{}iten}}, \bibinfo {author} {\bibfnamefont {J.}~\bibnamefont {Madsen}},
  \bibinfo {author} {\bibfnamefont {V.}~\bibnamefont {Naulin}}, \bibinfo
  {author} {\bibfnamefont {A.~H.}\ \bibnamefont {Nielsen}},\ and\ \bibinfo
  {author} {\bibfnamefont {J.~J.}\ \bibnamefont {Rasmussen}},\ }\href
  {https://doi.org/10.1063/1.5019662} {\bibfield  {journal} {\bibinfo
  {journal} {Physics of Plasmas}\ }\textbf {\bibinfo {volume} {25}},\ \bibinfo
  {pages} {032307} (\bibinfo {year} {2018})}\BibitemShut {NoStop}%
\bibitem [{\citenamefont {Nespoli}\ \emph {et~al.}(2017)\citenamefont
  {Nespoli}, \citenamefont {Furno}, \citenamefont {Halpern}, \citenamefont
  {Labit}, \citenamefont {Loizu}, \citenamefont {Ricci},\ and\ \citenamefont
  {Riva}}]{nespoli_non-linear_2017}%
  \BibitemOpen
  \bibfield  {author} {\bibinfo {author} {\bibfnamefont {F.}~\bibnamefont
  {Nespoli}}, \bibinfo {author} {\bibfnamefont {I.}~\bibnamefont {Furno}},
  \bibinfo {author} {\bibfnamefont {F.~D.}\ \bibnamefont {Halpern}}, \bibinfo
  {author} {\bibfnamefont {B.}~\bibnamefont {Labit}}, \bibinfo {author}
  {\bibfnamefont {J.}~\bibnamefont {Loizu}}, \bibinfo {author} {\bibfnamefont
  {P.}~\bibnamefont {Ricci}},\ and\ \bibinfo {author} {\bibfnamefont
  {F.}~\bibnamefont {Riva}},\ }\href
  {https://doi.org/https://doi.org/10.1016/j.nme.2016.10.019} {\bibfield
  {journal} {\bibinfo  {journal} {Nuclear Materials and Energy}\ }\textbf
  {\bibinfo {volume} {12}},\ \bibinfo {pages} {1205 } (\bibinfo {year}
  {2017})}\BibitemShut {NoStop}%
\bibitem [{\citenamefont {Biel}\ \emph {et~al.}(2019)\citenamefont {Biel},
  \citenamefont {Albanese}, \citenamefont {Ambrosino}, \citenamefont {Ariola},
  \citenamefont {Berkel}, \citenamefont {Bolshakova}, \citenamefont {Brunner},
  \citenamefont {Cavazzana}, \citenamefont {Cecconello}, \citenamefont
  {Conroy}, \citenamefont {Dinklage}, \citenamefont {Duran}, \citenamefont
  {Dux}, \citenamefont {Eade}, \citenamefont {Entler}, \citenamefont
  {Ericsson}, \citenamefont {Fable}, \citenamefont {Farina}, \citenamefont
  {Figini}, \citenamefont {Finotti}, \citenamefont {Franke}, \citenamefont
  {Giacomelli}, \citenamefont {Giannone}, \citenamefont {Gonzalez},
  \citenamefont {Hjalmarsson}, \citenamefont {Hron}, \citenamefont {Janky},
  \citenamefont {Kallenbach}, \citenamefont {Kogoj}, \citenamefont {K\"{o}nig},
  \citenamefont {Kudlacek}, \citenamefont {Luis}, \citenamefont {Malaquias},
  \citenamefont {Marchuk}, \citenamefont {Marchiori}, \citenamefont {Mattei},
  \citenamefont {Maviglia}, \citenamefont {Masi}, \citenamefont {Mazon},
  \citenamefont {Meister}, \citenamefont {Meyer}, \citenamefont {Micheletti},
  \citenamefont {Nowak}, \citenamefont {Piron}, \citenamefont {Pironti},
  \citenamefont {Rispoli}, \citenamefont {Rohde}, \citenamefont {Sergienko},
  \citenamefont {Shawish}, \citenamefont {Siccinio}, \citenamefont {Silva},
  \citenamefont {Silva}, \citenamefont {Sozzi}, \citenamefont {Tardocchi},
  \citenamefont {Tokar}, \citenamefont {Treutterer},\ and\ \citenamefont
  {Zohm}}]{biel_diagnostics_2019}%
  \BibitemOpen
  \bibfield  {author} {\bibinfo {author} {\bibfnamefont {W.}~\bibnamefont
  {Biel}}, \bibinfo {author} {\bibfnamefont {R.}~\bibnamefont {Albanese}},
  \bibinfo {author} {\bibfnamefont {R.}~\bibnamefont {Ambrosino}}, \bibinfo
  {author} {\bibfnamefont {M.}~\bibnamefont {Ariola}}, \bibinfo {author}
  {\bibfnamefont {M.~V.}\ \bibnamefont {Berkel}}, \bibinfo {author}
  {\bibfnamefont {I.}~\bibnamefont {Bolshakova}}, \bibinfo {author}
  {\bibfnamefont {K.~J.}\ \bibnamefont {Brunner}}, \bibinfo {author}
  {\bibfnamefont {R.}~\bibnamefont {Cavazzana}}, \bibinfo {author}
  {\bibfnamefont {M.}~\bibnamefont {Cecconello}}, \bibinfo {author}
  {\bibfnamefont {S.}~\bibnamefont {Conroy}}, \bibinfo {author} {\bibfnamefont
  {A.}~\bibnamefont {Dinklage}}, \bibinfo {author} {\bibfnamefont
  {I.}~\bibnamefont {Duran}}, \bibinfo {author} {\bibfnamefont
  {R.}~\bibnamefont {Dux}}, \bibinfo {author} {\bibfnamefont {T.}~\bibnamefont
  {Eade}}, \bibinfo {author} {\bibfnamefont {S.}~\bibnamefont {Entler}},
  \bibinfo {author} {\bibfnamefont {G.}~\bibnamefont {Ericsson}}, \bibinfo
  {author} {\bibfnamefont {E.}~\bibnamefont {Fable}}, \bibinfo {author}
  {\bibfnamefont {D.}~\bibnamefont {Farina}}, \bibinfo {author} {\bibfnamefont
  {L.}~\bibnamefont {Figini}}, \bibinfo {author} {\bibfnamefont
  {C.}~\bibnamefont {Finotti}}, \bibinfo {author} {\bibfnamefont
  {T.}~\bibnamefont {Franke}}, \bibinfo {author} {\bibfnamefont
  {L.}~\bibnamefont {Giacomelli}}, \bibinfo {author} {\bibfnamefont
  {L.}~\bibnamefont {Giannone}}, \bibinfo {author} {\bibfnamefont
  {W.}~\bibnamefont {Gonzalez}}, \bibinfo {author} {\bibfnamefont
  {A.}~\bibnamefont {Hjalmarsson}}, \bibinfo {author} {\bibfnamefont
  {M.}~\bibnamefont {Hron}}, \bibinfo {author} {\bibfnamefont {F.}~\bibnamefont
  {Janky}}, \bibinfo {author} {\bibfnamefont {A.}~\bibnamefont {Kallenbach}},
  \bibinfo {author} {\bibfnamefont {J.}~\bibnamefont {Kogoj}}, \bibinfo
  {author} {\bibfnamefont {R.}~\bibnamefont {K\"{o}nig}}, \bibinfo {author}
  {\bibfnamefont {O.}~\bibnamefont {Kudlacek}}, \bibinfo {author}
  {\bibfnamefont {R.}~\bibnamefont {Luis}}, \bibinfo {author} {\bibfnamefont
  {A.}~\bibnamefont {Malaquias}}, \bibinfo {author} {\bibfnamefont
  {O.}~\bibnamefont {Marchuk}}, \bibinfo {author} {\bibfnamefont
  {G.}~\bibnamefont {Marchiori}}, \bibinfo {author} {\bibfnamefont
  {M.}~\bibnamefont {Mattei}}, \bibinfo {author} {\bibfnamefont
  {F.}~\bibnamefont {Maviglia}}, \bibinfo {author} {\bibfnamefont {G.~D.}\
  \bibnamefont {Masi}}, \bibinfo {author} {\bibfnamefont {D.}~\bibnamefont
  {Mazon}}, \bibinfo {author} {\bibfnamefont {H.}~\bibnamefont {Meister}},
  \bibinfo {author} {\bibfnamefont {K.}~\bibnamefont {Meyer}}, \bibinfo
  {author} {\bibfnamefont {D.}~\bibnamefont {Micheletti}}, \bibinfo {author}
  {\bibfnamefont {S.}~\bibnamefont {Nowak}}, \bibinfo {author} {\bibfnamefont
  {C.}~\bibnamefont {Piron}}, \bibinfo {author} {\bibfnamefont
  {A.}~\bibnamefont {Pironti}}, \bibinfo {author} {\bibfnamefont
  {N.}~\bibnamefont {Rispoli}}, \bibinfo {author} {\bibfnamefont
  {V.}~\bibnamefont {Rohde}}, \bibinfo {author} {\bibfnamefont
  {G.}~\bibnamefont {Sergienko}}, \bibinfo {author} {\bibfnamefont {S.~E.}\
  \bibnamefont {Shawish}}, \bibinfo {author} {\bibfnamefont {M.}~\bibnamefont
  {Siccinio}}, \bibinfo {author} {\bibfnamefont {A.}~\bibnamefont {Silva}},
  \bibinfo {author} {\bibfnamefont {F.~d.}\ \bibnamefont {Silva}}, \bibinfo
  {author} {\bibfnamefont {C.}~\bibnamefont {Sozzi}}, \bibinfo {author}
  {\bibfnamefont {M.}~\bibnamefont {Tardocchi}}, \bibinfo {author}
  {\bibfnamefont {M.}~\bibnamefont {Tokar}}, \bibinfo {author} {\bibfnamefont
  {W.}~\bibnamefont {Treutterer}},\ and\ \bibinfo {author} {\bibfnamefont
  {H.}~\bibnamefont {Zohm}},\ }\href
  {https://doi.org/https://doi.org/10.1016/j.fusengdes.2018.12.092} {\bibfield
  {journal} {\bibinfo  {journal} {Fusion Engineering and Design}\ }\textbf
  {\bibinfo {volume} {146}},\ \bibinfo {pages} {465 } (\bibinfo {year}
  {2019})}\BibitemShut {NoStop}%
\bibitem [{\citenamefont {Griener}\ \emph {et~al.}(2020)\citenamefont
  {Griener}, \citenamefont {Wolfrum}, \citenamefont {Birkenmeier},
  \citenamefont {Faitsch}, \citenamefont {Fischer}, \citenamefont {Fuchert},
  \citenamefont {Gil}, \citenamefont {Harrer}, \citenamefont {Manz},
  \citenamefont {Wendler},\ and\ \citenamefont
  {Stroth}}]{griener_continuous_2020}%
  \BibitemOpen
  \bibfield  {author} {\bibinfo {author} {\bibfnamefont {M.}~\bibnamefont
  {Griener}}, \bibinfo {author} {\bibfnamefont {E.}~\bibnamefont {Wolfrum}},
  \bibinfo {author} {\bibfnamefont {G.}~\bibnamefont {Birkenmeier}}, \bibinfo
  {author} {\bibfnamefont {M.}~\bibnamefont {Faitsch}}, \bibinfo {author}
  {\bibfnamefont {R.}~\bibnamefont {Fischer}}, \bibinfo {author} {\bibfnamefont
  {G.}~\bibnamefont {Fuchert}}, \bibinfo {author} {\bibfnamefont
  {L.}~\bibnamefont {Gil}}, \bibinfo {author} {\bibfnamefont {G.~F.}\
  \bibnamefont {Harrer}}, \bibinfo {author} {\bibfnamefont {P.}~\bibnamefont
  {Manz}}, \bibinfo {author} {\bibfnamefont {D.}~\bibnamefont {Wendler}},\ and\
  \bibinfo {author} {\bibfnamefont {U.}~\bibnamefont {Stroth}},\ }\href
  {https://doi.org/https://doi.org/10.1016/j.nme.2020.100854} {\bibfield
  {journal} {\bibinfo  {journal} {Nuclear Materials and Energy}\ }\textbf
  {\bibinfo {volume} {25}},\ \bibinfo {pages} {100854} (\bibinfo {year}
  {2020})}\BibitemShut {NoStop}%
\bibitem [{\citenamefont {Bowman}\ \emph {et~al.}(2020)\citenamefont {Bowman},
  \citenamefont {Harrison}, \citenamefont {Lipschultz}, \citenamefont
  {Orchard}, \citenamefont {Gibson}, \citenamefont {Carr}, \citenamefont
  {Verhaegh},\ and\ \citenamefont {Myatra}}]{Bowman_2020}%
  \BibitemOpen
  \bibfield  {author} {\bibinfo {author} {\bibfnamefont {C.}~\bibnamefont
  {Bowman}}, \bibinfo {author} {\bibfnamefont {J.~R.}\ \bibnamefont
  {Harrison}}, \bibinfo {author} {\bibfnamefont {B.}~\bibnamefont
  {Lipschultz}}, \bibinfo {author} {\bibfnamefont {S.}~\bibnamefont {Orchard}},
  \bibinfo {author} {\bibfnamefont {K.~J.}\ \bibnamefont {Gibson}}, \bibinfo
  {author} {\bibfnamefont {M.}~\bibnamefont {Carr}}, \bibinfo {author}
  {\bibfnamefont {K.}~\bibnamefont {Verhaegh}},\ and\ \bibinfo {author}
  {\bibfnamefont {O.}~\bibnamefont {Myatra}},\ }\href
  {https://doi.org/10.1088/1361-6587/ab759b} {\bibfield  {journal} {\bibinfo
  {journal} {Plasma Physics and Controlled Fusion}\ }\textbf {\bibinfo {volume}
  {62}},\ \bibinfo {pages} {045014} (\bibinfo {year} {2020})}\BibitemShut
  {NoStop}%
\bibitem [{\citenamefont {Mathews}\ and\ \citenamefont
  {Hughes}(2021)}]{Mathews2020}%
  \BibitemOpen
  \bibfield  {author} {\bibinfo {author} {\bibfnamefont {A.}~\bibnamefont
  {Mathews}}\ and\ \bibinfo {author} {\bibfnamefont {J.}~\bibnamefont
  {Hughes}},\ }\href@noop {} {\bibinfo {title} {Quantifying experimental edge
  plasma evolution via multidimensional adaptive gaussian process regression}}
  (\bibinfo {year} {2021}),\ \Eprint {https://arxiv.org/abs/2103.01305}
  {arXiv:2103.01305 [physics.plasm-ph]} \BibitemShut {NoStop}%
\bibitem [{\citenamefont {Mathews}\ \emph
  {et~al.}(2021{\natexlab{a}})\citenamefont {Mathews}, \citenamefont {Hughes},
  \citenamefont {Terry}, \citenamefont {Goto}, \citenamefont {Stotler},
  \citenamefont {Reiter}, \citenamefont {Baek},\ and\ \citenamefont
  {Zweben}}]{Mathews2021}%
  \BibitemOpen
  \bibfield  {author} {\bibinfo {author} {\bibfnamefont {A.}~\bibnamefont
  {Mathews}}, \bibinfo {author} {\bibfnamefont {J.~W.}\ \bibnamefont {Hughes}},
  \bibinfo {author} {\bibfnamefont {J.}~\bibnamefont {Terry}}, \bibinfo
  {author} {\bibfnamefont {M.}~\bibnamefont {Goto}}, \bibinfo {author}
  {\bibfnamefont {D.}~\bibnamefont {Stotler}}, \bibinfo {author} {\bibfnamefont
  {D.}~\bibnamefont {Reiter}}, \bibinfo {author} {\bibfnamefont {S.-G.}\
  \bibnamefont {Baek}},\ and\ \bibinfo {author} {\bibfnamefont
  {S.}~\bibnamefont {Zweben}},\ }\href@noop {} {\bibinfo {title} {Deep
  collisional-radiative modelling of edge turbulent electron density and
  temperature fluctuations}} (\bibinfo {year} {2021}{\natexlab{a}}),\ \bibinfo
  {note} {in preparation}\BibitemShut {NoStop}%
\bibitem [{\citenamefont {Simakov}\ and\ \citenamefont
  {Catto}(2003)}]{Simakov-paper}%
  \BibitemOpen
  \bibfield  {author} {\bibinfo {author} {\bibfnamefont {A.~N.}\ \bibnamefont
  {Simakov}}\ and\ \bibinfo {author} {\bibfnamefont {P.~J.}\ \bibnamefont
  {Catto}},\ }\href {https://doi.org/10.1063/1.1623492} {\bibfield  {journal}
  {\bibinfo  {journal} {Physics of Plasmas}\ }\textbf {\bibinfo {volume}
  {10}},\ \bibinfo {pages} {4744} (\bibinfo {year} {2003})}\BibitemShut
  {NoStop}%
\bibitem [{\citenamefont {Wang}\ \emph
  {et~al.}(2020{\natexlab{a}})\citenamefont {Wang}, \citenamefont {Guo},
  \citenamefont {Xu}, \citenamefont {Leonard}, \citenamefont {Wu},
  \citenamefont {Groth}, \citenamefont {Jaervinen}, \citenamefont {Watkins},
  \citenamefont {Osborne}, \citenamefont {Thomas}, \citenamefont {Eldon},
  \citenamefont {Stangeby}, \citenamefont {Turco}, \citenamefont {Xu},
  \citenamefont {Wang}, \citenamefont {Wang},\ and\ \citenamefont
  {Liu}}]{osti_1630104}%
  \BibitemOpen
  \bibfield  {author} {\bibinfo {author} {\bibfnamefont {H.~Q.}\ \bibnamefont
  {Wang}}, \bibinfo {author} {\bibfnamefont {H.~Y.}\ \bibnamefont {Guo}},
  \bibinfo {author} {\bibfnamefont {G.~S.}\ \bibnamefont {Xu}}, \bibinfo
  {author} {\bibfnamefont {A.~W.}\ \bibnamefont {Leonard}}, \bibinfo {author}
  {\bibfnamefont {X.~Q.}\ \bibnamefont {Wu}}, \bibinfo {author} {\bibfnamefont
  {M.}~\bibnamefont {Groth}}, \bibinfo {author} {\bibfnamefont {A.~E.}\
  \bibnamefont {Jaervinen}}, \bibinfo {author} {\bibfnamefont {J.~G.}\
  \bibnamefont {Watkins}}, \bibinfo {author} {\bibfnamefont {T.~H.}\
  \bibnamefont {Osborne}}, \bibinfo {author} {\bibfnamefont {D.~M.}\
  \bibnamefont {Thomas}}, \bibinfo {author} {\bibfnamefont {D.}~\bibnamefont
  {Eldon}}, \bibinfo {author} {\bibfnamefont {P.~C.}\ \bibnamefont {Stangeby}},
  \bibinfo {author} {\bibfnamefont {F.}~\bibnamefont {Turco}}, \bibinfo
  {author} {\bibfnamefont {J.~C.}\ \bibnamefont {Xu}}, \bibinfo {author}
  {\bibfnamefont {L.}~\bibnamefont {Wang}}, \bibinfo {author} {\bibfnamefont
  {Y.~F.}\ \bibnamefont {Wang}},\ and\ \bibinfo {author} {\bibfnamefont
  {J.~B.}\ \bibnamefont {Liu}},\ }\href@noop {} {\bibfield  {journal} {\bibinfo
   {journal} {Physical Review Letters}\ }\textbf {\bibinfo {volume} {124}}
  (\bibinfo {year} {2020}{\natexlab{a}})}\BibitemShut {NoStop}%
\bibitem [{\citenamefont {Zhang}\ \emph {et~al.}(2020)\citenamefont {Zhang},
  \citenamefont {Guo},\ and\ \citenamefont {Diamond}}]{osti_1762124}%
  \BibitemOpen
  \bibfield  {author} {\bibinfo {author} {\bibfnamefont {Y.}~\bibnamefont
  {Zhang}}, \bibinfo {author} {\bibfnamefont {Z.~B.}\ \bibnamefont {Guo}},\
  and\ \bibinfo {author} {\bibfnamefont {P.~H.}\ \bibnamefont {Diamond}},\
  }\href@noop {} {\bibfield  {journal} {\bibinfo  {journal} {Physical Review
  Letters}\ }\textbf {\bibinfo {volume} {125}} (\bibinfo {year}
  {2020})}\BibitemShut {NoStop}%
\bibitem [{\citenamefont {Kuang}\ \emph {et~al.}(2020)\citenamefont {Kuang},
  \citenamefont {Ballinger}, \citenamefont {Brunner}, \citenamefont {Canik},
  \citenamefont {Creely}, \citenamefont {Gray}, \citenamefont {Greenwald},
  \citenamefont {Hughes}, \citenamefont {Irby}, \citenamefont {LaBombard},\
  and\ \citenamefont {et~al.}}]{kuang_SPARC}%
  \BibitemOpen
  \bibfield  {author} {\bibinfo {author} {\bibfnamefont {A.~Q.}\ \bibnamefont
  {Kuang}}, \bibinfo {author} {\bibfnamefont {S.}~\bibnamefont {Ballinger}},
  \bibinfo {author} {\bibfnamefont {D.}~\bibnamefont {Brunner}}, \bibinfo
  {author} {\bibfnamefont {J.}~\bibnamefont {Canik}}, \bibinfo {author}
  {\bibfnamefont {A.~J.}\ \bibnamefont {Creely}}, \bibinfo {author}
  {\bibfnamefont {T.}~\bibnamefont {Gray}}, \bibinfo {author} {\bibfnamefont
  {M.}~\bibnamefont {Greenwald}}, \bibinfo {author} {\bibfnamefont {J.~W.}\
  \bibnamefont {Hughes}}, \bibinfo {author} {\bibfnamefont {J.}~\bibnamefont
  {Irby}}, \bibinfo {author} {\bibfnamefont {B.}~\bibnamefont {LaBombard}},\
  and\ \bibinfo {author} {\bibnamefont {et~al.}},\ }\href
  {https://doi.org/10.1017/S0022377820001117} {\bibfield  {journal} {\bibinfo
  {journal} {Journal of Plasma Physics}\ }\textbf {\bibinfo {volume} {86}},\
  \bibinfo {pages} {865860505} (\bibinfo {year} {2020})}\BibitemShut {NoStop}%
\bibitem [{\citenamefont {Kuang}\ \emph {et~al.}(2018)\citenamefont {Kuang},
  \citenamefont {Cao}, \citenamefont {Creely}, \citenamefont {Dennett},
  \citenamefont {Hecla}, \citenamefont {LaBombard}, \citenamefont {Tinguely},
  \citenamefont {Tolman}, \citenamefont {Hoffman}, \citenamefont {Major},
  \citenamefont {Ruiz}, \citenamefont {Brunner}, \citenamefont {Grover},
  \citenamefont {Laughman}, \citenamefont {Sorbom},\ and\ \citenamefont
  {Whyte}}]{kuang_ARC}%
  \BibitemOpen
  \bibfield  {author} {\bibinfo {author} {\bibfnamefont {A.~Q.}\ \bibnamefont
  {Kuang}}, \bibinfo {author} {\bibfnamefont {N.~M.}\ \bibnamefont {Cao}},
  \bibinfo {author} {\bibfnamefont {A.~J.}\ \bibnamefont {Creely}}, \bibinfo
  {author} {\bibfnamefont {C.~A.}\ \bibnamefont {Dennett}}, \bibinfo {author}
  {\bibfnamefont {J.}~\bibnamefont {Hecla}}, \bibinfo {author} {\bibfnamefont
  {B.}~\bibnamefont {LaBombard}}, \bibinfo {author} {\bibfnamefont {R.~A.}\
  \bibnamefont {Tinguely}}, \bibinfo {author} {\bibfnamefont {E.~A.}\
  \bibnamefont {Tolman}}, \bibinfo {author} {\bibfnamefont {H.}~\bibnamefont
  {Hoffman}}, \bibinfo {author} {\bibfnamefont {M.}~\bibnamefont {Major}},
  \bibinfo {author} {\bibfnamefont {J.~R.}\ \bibnamefont {Ruiz}}, \bibinfo
  {author} {\bibfnamefont {D.}~\bibnamefont {Brunner}}, \bibinfo {author}
  {\bibfnamefont {P.}~\bibnamefont {Grover}}, \bibinfo {author} {\bibfnamefont
  {C.}~\bibnamefont {Laughman}}, \bibinfo {author} {\bibfnamefont {B.~N.}\
  \bibnamefont {Sorbom}},\ and\ \bibinfo {author} {\bibfnamefont {D.~G.}\
  \bibnamefont {Whyte}},\ }\href
  {https://doi.org/https://doi.org/10.1016/j.fusengdes.2018.09.007} {\bibfield
  {journal} {\bibinfo  {journal} {Fusion Engineering and Design}\ }\textbf
  {\bibinfo {volume} {137}},\ \bibinfo {pages} {221 } (\bibinfo {year}
  {2018})}\BibitemShut {NoStop}%
\bibitem [{\citenamefont {Lagaris}\ \emph {et~al.}(1998)\citenamefont
  {Lagaris}, \citenamefont {Likas},\ and\ \citenamefont
  {Fotiadis}}]{Lagaris_1998}%
  \BibitemOpen
  \bibfield  {author} {\bibinfo {author} {\bibfnamefont {I.}~\bibnamefont
  {Lagaris}}, \bibinfo {author} {\bibfnamefont {A.}~\bibnamefont {Likas}},\
  and\ \bibinfo {author} {\bibfnamefont {D.}~\bibnamefont {Fotiadis}},\ }\href
  {https://doi.org/10.1109/72.712178} {\bibfield  {journal} {\bibinfo
  {journal} {IEEE Transactions on Neural Networks}\ }\textbf {\bibinfo {volume}
  {9}},\ \bibinfo {pages} {987–1000} (\bibinfo {year} {1998})}\BibitemShut
  {NoStop}%
\bibitem [{\citenamefont {Raissi}\ \emph {et~al.}(2019)\citenamefont {Raissi},
  \citenamefont {Perdikaris},\ and\ \citenamefont
  {Karniadakis}}]{raissi2017physics}%
  \BibitemOpen
  \bibfield  {author} {\bibinfo {author} {\bibfnamefont {M.}~\bibnamefont
  {Raissi}}, \bibinfo {author} {\bibfnamefont {P.}~\bibnamefont {Perdikaris}},\
  and\ \bibinfo {author} {\bibfnamefont {G.}~\bibnamefont {Karniadakis}},\
  }\href {https://doi.org/https://doi.org/10.1016/j.jcp.2018.10.045} {\bibfield
   {journal} {\bibinfo  {journal} {Journal of Computational Physics}\ }\textbf
  {\bibinfo {volume} {378}},\ \bibinfo {pages} {686} (\bibinfo {year}
  {2019})}\BibitemShut {NoStop}%
\bibitem [{\citenamefont {Michoski}\ \emph {et~al.}(2020)\citenamefont
  {Michoski}, \citenamefont {Milosavljevi\'c}, \citenamefont {Oliver},\ and\
  \citenamefont {Hatch}}]{MICHOSKI2020193}%
  \BibitemOpen
  \bibfield  {author} {\bibinfo {author} {\bibfnamefont {C.}~\bibnamefont
  {Michoski}}, \bibinfo {author} {\bibfnamefont {M.}~\bibnamefont
  {Milosavljevi\'c}}, \bibinfo {author} {\bibfnamefont {T.}~\bibnamefont
  {Oliver}},\ and\ \bibinfo {author} {\bibfnamefont {D.~R.}\ \bibnamefont
  {Hatch}},\ }\href@noop {} {\bibfield  {journal} {\bibinfo  {journal}
  {Neurocomputing}\ }\textbf {\bibinfo {volume} {399}},\ \bibinfo {pages} {193
  } (\bibinfo {year} {2020})}\BibitemShut {NoStop}%
\bibitem [{\citenamefont {Sirignano}\ and\ \citenamefont
  {Spiliopoulos}(2018)}]{SIRIGNANO20181339}%
  \BibitemOpen
  \bibfield  {author} {\bibinfo {author} {\bibfnamefont {J.}~\bibnamefont
  {Sirignano}}\ and\ \bibinfo {author} {\bibfnamefont {K.}~\bibnamefont
  {Spiliopoulos}},\ }\href
  {https://doi.org/https://doi.org/10.1016/j.jcp.2018.08.029} {\bibfield
  {journal} {\bibinfo  {journal} {Journal of Computational Physics}\ }\textbf
  {\bibinfo {volume} {375}},\ \bibinfo {pages} {1339 } (\bibinfo {year}
  {2018})}\BibitemShut {NoStop}%
\bibitem [{\citenamefont {Zhu}\ \emph {et~al.}(2018)\citenamefont {Zhu},
  \citenamefont {Francisquez},\ and\ \citenamefont {Rogers}}]{osti_1488775}%
  \BibitemOpen
  \bibfield  {author} {\bibinfo {author} {\bibfnamefont {B.}~\bibnamefont
  {Zhu}}, \bibinfo {author} {\bibfnamefont {M.}~\bibnamefont {Francisquez}},\
  and\ \bibinfo {author} {\bibfnamefont {B.~N.}\ \bibnamefont {Rogers}},\
  }\href@noop {} {\bibfield  {journal} {\bibinfo  {journal} {Computer Physics
  Communications}\ }\textbf {\bibinfo {volume} {232}} (\bibinfo {year}
  {2018})}\BibitemShut {NoStop}%
\bibitem [{\citenamefont {Francisquez}\ \emph {et~al.}(2020)\citenamefont
  {Francisquez}, \citenamefont {Bernard}, \citenamefont {Zhu}, \citenamefont
  {Hakim}, \citenamefont {Rogers},\ and\ \citenamefont
  {Hammett}}]{francisquez2020fluid}%
  \BibitemOpen
  \bibfield  {author} {\bibinfo {author} {\bibfnamefont {M.}~\bibnamefont
  {Francisquez}}, \bibinfo {author} {\bibfnamefont {T.~N.}\ \bibnamefont
  {Bernard}}, \bibinfo {author} {\bibfnamefont {B.}~\bibnamefont {Zhu}},
  \bibinfo {author} {\bibfnamefont {A.}~\bibnamefont {Hakim}}, \bibinfo
  {author} {\bibfnamefont {B.~N.}\ \bibnamefont {Rogers}},\ and\ \bibinfo
  {author} {\bibfnamefont {G.~W.}\ \bibnamefont {Hammett}},\ }\href
  {https://doi.org/10.1063/5.0005333} {\bibfield  {journal} {\bibinfo
  {journal} {Physics of Plasmas}\ }\textbf {\bibinfo {volume} {27}},\ \bibinfo
  {pages} {082301} (\bibinfo {year} {2020})}\BibitemShut {NoStop}%
\bibitem [{\citenamefont {Belli}\ and\ \citenamefont
  {Candy}(2008)}]{Belli_2008}%
  \BibitemOpen
  \bibfield  {author} {\bibinfo {author} {\bibfnamefont {E.~A.}\ \bibnamefont
  {Belli}}\ and\ \bibinfo {author} {\bibfnamefont {J.}~\bibnamefont {Candy}},\
  }\href {https://doi.org/10.1088/0741-3335/50/9/095010} {\bibfield  {journal}
  {\bibinfo  {journal} {Plasma Physics and Controlled Fusion}\ }\textbf
  {\bibinfo {volume} {50}},\ \bibinfo {pages} {095010} (\bibinfo {year}
  {2008})}\BibitemShut {NoStop}%
\bibitem [{\citenamefont {Madsen}(2013)}]{Full-F}%
  \BibitemOpen
  \bibfield  {author} {\bibinfo {author} {\bibfnamefont {J.}~\bibnamefont
  {Madsen}},\ }\href {https://doi.org/10.1063/1.4813241} {\bibfield  {journal}
  {\bibinfo  {journal} {Physics of Plasmas}\ }\textbf {\bibinfo {volume}
  {20}},\ \bibinfo {pages} {072301} (\bibinfo {year} {2013})}\BibitemShut
  {NoStop}%
\bibitem [{\citenamefont {Held}\ \emph {et~al.}(2020)\citenamefont {Held},
  \citenamefont {Wiesenberger},\ and\ \citenamefont {Kendl}}]{Held_2020}%
  \BibitemOpen
  \bibfield  {author} {\bibinfo {author} {\bibfnamefont {M.}~\bibnamefont
  {Held}}, \bibinfo {author} {\bibfnamefont {M.}~\bibnamefont {Wiesenberger}},\
  and\ \bibinfo {author} {\bibfnamefont {A.}~\bibnamefont {Kendl}},\ }\href
  {https://doi.org/10.1088/1741-4326/ab7a68} {\bibfield  {journal} {\bibinfo
  {journal} {Nuclear Fusion}\ }\textbf {\bibinfo {volume} {60}},\ \bibinfo
  {pages} {066014} (\bibinfo {year} {2020})}\BibitemShut {NoStop}%
\bibitem [{\citenamefont {Garcia}\ \emph {et~al.}(2013)\citenamefont {Garcia},
  \citenamefont {Fritzner}, \citenamefont {Kube}, \citenamefont {Cziegler},
  \citenamefont {LaBombard},\ and\ \citenamefont
  {Terry}}]{Garcia_SOL_fluctuations}%
  \BibitemOpen
  \bibfield  {author} {\bibinfo {author} {\bibfnamefont {O.~E.}\ \bibnamefont
  {Garcia}}, \bibinfo {author} {\bibfnamefont {S.~M.}\ \bibnamefont
  {Fritzner}}, \bibinfo {author} {\bibfnamefont {R.}~\bibnamefont {Kube}},
  \bibinfo {author} {\bibfnamefont {I.}~\bibnamefont {Cziegler}}, \bibinfo
  {author} {\bibfnamefont {B.}~\bibnamefont {LaBombard}},\ and\ \bibinfo
  {author} {\bibfnamefont {J.~L.}\ \bibnamefont {Terry}},\ }\href
  {https://doi.org/10.1063/1.4802942} {\bibfield  {journal} {\bibinfo
  {journal} {Physics of Plasmas}\ }\textbf {\bibinfo {volume} {20}},\ \bibinfo
  {pages} {055901} (\bibinfo {year} {2013})}\BibitemShut {NoStop}%
\bibitem [{\citenamefont {Scott}(2003)}]{DRB_consistency3}%
  \BibitemOpen
  \bibfield  {author} {\bibinfo {author} {\bibfnamefont {B.}~\bibnamefont
  {Scott}},\ }\href {https://doi.org/10.1063/1.1545100} {\bibfield  {journal}
  {\bibinfo  {journal} {Physics of Plasmas}\ }\textbf {\bibinfo {volume}
  {10}},\ \bibinfo {pages} {963} (\bibinfo {year} {2003})}\BibitemShut
  {NoStop}%
\bibitem [{\citenamefont {Zholobenko}\ \emph {et~al.}(2021)\citenamefont
  {Zholobenko}, \citenamefont {Body}, \citenamefont {Manz}, \citenamefont
  {Stegmeir}, \citenamefont {Zhu}, \citenamefont {Griener}, \citenamefont
  {Conway}, \citenamefont {Coster},\ and\ \citenamefont
  {Jenko}}]{Zholobenko_2021}%
  \BibitemOpen
  \bibfield  {author} {\bibinfo {author} {\bibfnamefont {W.}~\bibnamefont
  {Zholobenko}}, \bibinfo {author} {\bibfnamefont {T.~A.}\ \bibnamefont
  {Body}}, \bibinfo {author} {\bibfnamefont {P.}~\bibnamefont {Manz}}, \bibinfo
  {author} {\bibfnamefont {A.}~\bibnamefont {Stegmeir}}, \bibinfo {author}
  {\bibfnamefont {B.}~\bibnamefont {Zhu}}, \bibinfo {author} {\bibfnamefont
  {M.}~\bibnamefont {Griener}}, \bibinfo {author} {\bibfnamefont {G.~D.}\
  \bibnamefont {Conway}}, \bibinfo {author} {\bibfnamefont {D.~P.}\
  \bibnamefont {Coster}},\ and\ \bibinfo {author} {\bibfnamefont
  {F.}~\bibnamefont {Jenko}},\ }\href@noop {} {\bibfield  {journal} {\bibinfo
  {journal} {Plasma Physics and Controlled Fusion}\ } (\bibinfo {year}
  {2021})}\BibitemShut {NoStop}%
\bibitem [{\citenamefont {Greenwald}\ \emph {et~al.}(2014)\citenamefont
  {Greenwald}, \citenamefont {Bader}, \citenamefont {Baek}, \citenamefont
  {Bakhtiari}, \citenamefont {Barnard}, \citenamefont {Beck}, \citenamefont
  {Bergerson}, \citenamefont {Bespamyatnov}, \citenamefont {Bonoli},
  \citenamefont {Brower}, \citenamefont {Brunner}, \citenamefont {Burke},
  \citenamefont {Candy}, \citenamefont {Churchill}, \citenamefont {Cziegler},
  \citenamefont {Diallo}, \citenamefont {Dominguez}, \citenamefont {Duval},
  \citenamefont {Edlund}, \citenamefont {Ennever}, \citenamefont {Ernst},
  \citenamefont {Faust}, \citenamefont {Fiore}, \citenamefont {Fredian},
  \citenamefont {Garcia}, \citenamefont {Gao}, \citenamefont {Goetz},
  \citenamefont {Golfinopoulos}, \citenamefont {Granetz}, \citenamefont
  {Grulke}, \citenamefont {Hartwig}, \citenamefont {Horne}, \citenamefont
  {Howard}, \citenamefont {Hubbard}, \citenamefont {Hughes}, \citenamefont
  {Hutchinson}, \citenamefont {Irby}, \citenamefont {Izzo}, \citenamefont
  {Kessel}, \citenamefont {LaBombard}, \citenamefont {Lau}, \citenamefont {Li},
  \citenamefont {Lin}, \citenamefont {Lipschultz}, \citenamefont {Loarte},
  \citenamefont {Marmar}, \citenamefont {Mazurenko}, \citenamefont {McCracken},
  \citenamefont {McDermott}, \citenamefont {Meneghini}, \citenamefont
  {Mikkelsen}, \citenamefont {Mossessian}, \citenamefont {Mumgaard},
  \citenamefont {Myra}, \citenamefont {Nelson-Melby}, \citenamefont {Ochoukov},
  \citenamefont {Olynyk}, \citenamefont {Parker}, \citenamefont {Pitcher},
  \citenamefont {Podpaly}, \citenamefont {Porkolab}, \citenamefont {Reinke},
  \citenamefont {Rice}, \citenamefont {Rowan}, \citenamefont {Schmidt},
  \citenamefont {Scott}, \citenamefont {Shiraiwa}, \citenamefont {Sierchio},
  \citenamefont {Smick}, \citenamefont {Snipes}, \citenamefont {Snyder},
  \citenamefont {Sorbom}, \citenamefont {Stillerman}, \citenamefont {Sung},
  \citenamefont {Takase}, \citenamefont {Tang}, \citenamefont {Terry},
  \citenamefont {Terry}, \citenamefont {Theiler}, \citenamefont
  {Tronchin-James}, \citenamefont {Tsujii}, \citenamefont {Vieira},
  \citenamefont {Walk}, \citenamefont {Wallace}, \citenamefont {White},
  \citenamefont {Whyte}, \citenamefont {Wilson}, \citenamefont {Wolfe},
  \citenamefont {Wright}, \citenamefont {Wright}, \citenamefont {Wukitch},\
  and\ \citenamefont {Zweben}}]{Alcator_Greenwald}%
  \BibitemOpen
  \bibfield  {author} {\bibinfo {author} {\bibfnamefont {M.}~\bibnamefont
  {Greenwald}}, \bibinfo {author} {\bibfnamefont {A.}~\bibnamefont {Bader}},
  \bibinfo {author} {\bibfnamefont {S.}~\bibnamefont {Baek}}, \bibinfo {author}
  {\bibfnamefont {M.}~\bibnamefont {Bakhtiari}}, \bibinfo {author}
  {\bibfnamefont {H.}~\bibnamefont {Barnard}}, \bibinfo {author} {\bibfnamefont
  {W.}~\bibnamefont {Beck}}, \bibinfo {author} {\bibfnamefont {W.}~\bibnamefont
  {Bergerson}}, \bibinfo {author} {\bibfnamefont {I.}~\bibnamefont
  {Bespamyatnov}}, \bibinfo {author} {\bibfnamefont {P.}~\bibnamefont
  {Bonoli}}, \bibinfo {author} {\bibfnamefont {D.}~\bibnamefont {Brower}},
  \bibinfo {author} {\bibfnamefont {D.}~\bibnamefont {Brunner}}, \bibinfo
  {author} {\bibfnamefont {W.}~\bibnamefont {Burke}}, \bibinfo {author}
  {\bibfnamefont {J.}~\bibnamefont {Candy}}, \bibinfo {author} {\bibfnamefont
  {M.}~\bibnamefont {Churchill}}, \bibinfo {author} {\bibfnamefont
  {I.}~\bibnamefont {Cziegler}}, \bibinfo {author} {\bibfnamefont
  {A.}~\bibnamefont {Diallo}}, \bibinfo {author} {\bibfnamefont
  {A.}~\bibnamefont {Dominguez}}, \bibinfo {author} {\bibfnamefont
  {B.}~\bibnamefont {Duval}}, \bibinfo {author} {\bibfnamefont
  {E.}~\bibnamefont {Edlund}}, \bibinfo {author} {\bibfnamefont
  {P.}~\bibnamefont {Ennever}}, \bibinfo {author} {\bibfnamefont
  {D.}~\bibnamefont {Ernst}}, \bibinfo {author} {\bibfnamefont
  {I.}~\bibnamefont {Faust}}, \bibinfo {author} {\bibfnamefont
  {C.}~\bibnamefont {Fiore}}, \bibinfo {author} {\bibfnamefont
  {T.}~\bibnamefont {Fredian}}, \bibinfo {author} {\bibfnamefont
  {O.}~\bibnamefont {Garcia}}, \bibinfo {author} {\bibfnamefont
  {C.}~\bibnamefont {Gao}}, \bibinfo {author} {\bibfnamefont {J.}~\bibnamefont
  {Goetz}}, \bibinfo {author} {\bibfnamefont {T.}~\bibnamefont
  {Golfinopoulos}}, \bibinfo {author} {\bibfnamefont {R.}~\bibnamefont
  {Granetz}}, \bibinfo {author} {\bibfnamefont {O.}~\bibnamefont {Grulke}},
  \bibinfo {author} {\bibfnamefont {Z.}~\bibnamefont {Hartwig}}, \bibinfo
  {author} {\bibfnamefont {S.}~\bibnamefont {Horne}}, \bibinfo {author}
  {\bibfnamefont {N.}~\bibnamefont {Howard}}, \bibinfo {author} {\bibfnamefont
  {A.}~\bibnamefont {Hubbard}}, \bibinfo {author} {\bibfnamefont
  {J.}~\bibnamefont {Hughes}}, \bibinfo {author} {\bibfnamefont
  {I.}~\bibnamefont {Hutchinson}}, \bibinfo {author} {\bibfnamefont
  {J.}~\bibnamefont {Irby}}, \bibinfo {author} {\bibfnamefont {V.}~\bibnamefont
  {Izzo}}, \bibinfo {author} {\bibfnamefont {C.}~\bibnamefont {Kessel}},
  \bibinfo {author} {\bibfnamefont {B.}~\bibnamefont {LaBombard}}, \bibinfo
  {author} {\bibfnamefont {C.}~\bibnamefont {Lau}}, \bibinfo {author}
  {\bibfnamefont {C.}~\bibnamefont {Li}}, \bibinfo {author} {\bibfnamefont
  {Y.}~\bibnamefont {Lin}}, \bibinfo {author} {\bibfnamefont {B.}~\bibnamefont
  {Lipschultz}}, \bibinfo {author} {\bibfnamefont {A.}~\bibnamefont {Loarte}},
  \bibinfo {author} {\bibfnamefont {E.}~\bibnamefont {Marmar}}, \bibinfo
  {author} {\bibfnamefont {A.}~\bibnamefont {Mazurenko}}, \bibinfo {author}
  {\bibfnamefont {G.}~\bibnamefont {McCracken}}, \bibinfo {author}
  {\bibfnamefont {R.}~\bibnamefont {McDermott}}, \bibinfo {author}
  {\bibfnamefont {O.}~\bibnamefont {Meneghini}}, \bibinfo {author}
  {\bibfnamefont {D.}~\bibnamefont {Mikkelsen}}, \bibinfo {author}
  {\bibfnamefont {D.}~\bibnamefont {Mossessian}}, \bibinfo {author}
  {\bibfnamefont {R.}~\bibnamefont {Mumgaard}}, \bibinfo {author}
  {\bibfnamefont {J.}~\bibnamefont {Myra}}, \bibinfo {author} {\bibfnamefont
  {E.}~\bibnamefont {Nelson-Melby}}, \bibinfo {author} {\bibfnamefont
  {R.}~\bibnamefont {Ochoukov}}, \bibinfo {author} {\bibfnamefont
  {G.}~\bibnamefont {Olynyk}}, \bibinfo {author} {\bibfnamefont
  {R.}~\bibnamefont {Parker}}, \bibinfo {author} {\bibfnamefont
  {S.}~\bibnamefont {Pitcher}}, \bibinfo {author} {\bibfnamefont
  {Y.}~\bibnamefont {Podpaly}}, \bibinfo {author} {\bibfnamefont
  {M.}~\bibnamefont {Porkolab}}, \bibinfo {author} {\bibfnamefont
  {M.}~\bibnamefont {Reinke}}, \bibinfo {author} {\bibfnamefont
  {J.}~\bibnamefont {Rice}}, \bibinfo {author} {\bibfnamefont {W.}~\bibnamefont
  {Rowan}}, \bibinfo {author} {\bibfnamefont {A.}~\bibnamefont {Schmidt}},
  \bibinfo {author} {\bibfnamefont {S.}~\bibnamefont {Scott}}, \bibinfo
  {author} {\bibfnamefont {S.}~\bibnamefont {Shiraiwa}}, \bibinfo {author}
  {\bibfnamefont {J.}~\bibnamefont {Sierchio}}, \bibinfo {author}
  {\bibfnamefont {N.}~\bibnamefont {Smick}}, \bibinfo {author} {\bibfnamefont
  {J.~A.}\ \bibnamefont {Snipes}}, \bibinfo {author} {\bibfnamefont
  {P.}~\bibnamefont {Snyder}}, \bibinfo {author} {\bibfnamefont
  {B.}~\bibnamefont {Sorbom}}, \bibinfo {author} {\bibfnamefont
  {J.}~\bibnamefont {Stillerman}}, \bibinfo {author} {\bibfnamefont
  {C.}~\bibnamefont {Sung}}, \bibinfo {author} {\bibfnamefont {Y.}~\bibnamefont
  {Takase}}, \bibinfo {author} {\bibfnamefont {V.}~\bibnamefont {Tang}},
  \bibinfo {author} {\bibfnamefont {J.}~\bibnamefont {Terry}}, \bibinfo
  {author} {\bibfnamefont {D.}~\bibnamefont {Terry}}, \bibinfo {author}
  {\bibfnamefont {C.}~\bibnamefont {Theiler}}, \bibinfo {author} {\bibfnamefont
  {A.}~\bibnamefont {Tronchin-James}}, \bibinfo {author} {\bibfnamefont
  {N.}~\bibnamefont {Tsujii}}, \bibinfo {author} {\bibfnamefont
  {R.}~\bibnamefont {Vieira}}, \bibinfo {author} {\bibfnamefont
  {J.}~\bibnamefont {Walk}}, \bibinfo {author} {\bibfnamefont {G.}~\bibnamefont
  {Wallace}}, \bibinfo {author} {\bibfnamefont {A.}~\bibnamefont {White}},
  \bibinfo {author} {\bibfnamefont {D.}~\bibnamefont {Whyte}}, \bibinfo
  {author} {\bibfnamefont {J.}~\bibnamefont {Wilson}}, \bibinfo {author}
  {\bibfnamefont {S.}~\bibnamefont {Wolfe}}, \bibinfo {author} {\bibfnamefont
  {G.}~\bibnamefont {Wright}}, \bibinfo {author} {\bibfnamefont
  {J.}~\bibnamefont {Wright}}, \bibinfo {author} {\bibfnamefont
  {S.}~\bibnamefont {Wukitch}},\ and\ \bibinfo {author} {\bibfnamefont
  {S.}~\bibnamefont {Zweben}},\ }\href {https://doi.org/10.1063/1.4901920}
  {\bibfield  {journal} {\bibinfo  {journal} {Physics of Plasmas}\ }\textbf
  {\bibinfo {volume} {21}},\ \bibinfo {pages} {110501} (\bibinfo {year}
  {2014})}\BibitemShut {NoStop}%
\bibitem [{\citenamefont {LaBombard}\ \emph {et~al.}(2005)\citenamefont
  {LaBombard}, \citenamefont {Hughes}, \citenamefont {Mossessian},
  \citenamefont {Greenwald}, \citenamefont {Lipschultz},\ and\ \citenamefont
  {Terry}}]{labombard_evidence_2005}%
  \BibitemOpen
  \bibfield  {author} {\bibinfo {author} {\bibfnamefont {B.}~\bibnamefont
  {LaBombard}}, \bibinfo {author} {\bibfnamefont {J.~W.}\ \bibnamefont
  {Hughes}}, \bibinfo {author} {\bibfnamefont {D.}~\bibnamefont {Mossessian}},
  \bibinfo {author} {\bibfnamefont {M.}~\bibnamefont {Greenwald}}, \bibinfo
  {author} {\bibfnamefont {B.}~\bibnamefont {Lipschultz}},\ and\ \bibinfo
  {author} {\bibfnamefont {J.~L.}\ \bibnamefont {Terry}},\ }\href
  {https://doi.org/10.1088/0029-5515/45/12/022} {\bibfield  {journal} {\bibinfo
   {journal} {Nuclear Fusion}\ }\textbf {\bibinfo {volume} {45}},\ \bibinfo
  {pages} {1658} (\bibinfo {year} {2005})},\ \bibinfo {note} {iOP
  Publishing}\BibitemShut {NoStop}%
\bibitem [{\citenamefont {Leddy}\ \emph {et~al.}(2015)\citenamefont {Leddy},
  \citenamefont {Dudson},\ and\ \citenamefont {and}}]{Leddy_full_velocity}%
  \BibitemOpen
  \bibfield  {author} {\bibinfo {author} {\bibfnamefont {J.}~\bibnamefont
  {Leddy}}, \bibinfo {author} {\bibfnamefont {B.}~\bibnamefont {Dudson}},\ and\
  \bibinfo {author} {\bibfnamefont {M.~R.}\ \bibnamefont {and}},\ }\href
  {https://doi.org/10.1088/0741-3335/57/12/125016} {\bibfield  {journal}
  {\bibinfo  {journal} {Plasma Physics and Controlled Fusion}\ }\textbf
  {\bibinfo {volume} {57}},\ \bibinfo {pages} {125016} (\bibinfo {year}
  {2015})}\BibitemShut {NoStop}%
\bibitem [{\citenamefont {Thrys\o{}e}\ \emph {et~al.}(2020)\citenamefont
  {Thrys\o{}e}, \citenamefont {Naulin}, \citenamefont {Nielsen},\ and\
  \citenamefont {Juul~Rasmussen}}]{Thrysoe_2020}%
  \BibitemOpen
  \bibfield  {author} {\bibinfo {author} {\bibfnamefont {A.~S.}\ \bibnamefont
  {Thrys\o{}e}}, \bibinfo {author} {\bibfnamefont {V.}~\bibnamefont {Naulin}},
  \bibinfo {author} {\bibfnamefont {A.~H.}\ \bibnamefont {Nielsen}},\ and\
  \bibinfo {author} {\bibfnamefont {J.}~\bibnamefont {Juul~Rasmussen}},\ }\href
  {https://doi.org/10.1063/5.0003262} {\bibfield  {journal} {\bibinfo
  {journal} {Physics of Plasmas}\ }\textbf {\bibinfo {volume} {27}},\ \bibinfo
  {pages} {052302} (\bibinfo {year} {2020})}\BibitemShut {NoStop}%
\bibitem [{\citenamefont {Stegmeir}\ \emph {et~al.}(2019)\citenamefont
  {Stegmeir}, \citenamefont {Ross}, \citenamefont {Body}, \citenamefont
  {Francisquez}, \citenamefont {Zholobenko}, \citenamefont {Coster},
  \citenamefont {Maj}, \citenamefont {Manz}, \citenamefont {Jenko},
  \citenamefont {Rogers},\ and\ \citenamefont {Kang}}]{GRILLIX_2018}%
  \BibitemOpen
  \bibfield  {author} {\bibinfo {author} {\bibfnamefont {A.}~\bibnamefont
  {Stegmeir}}, \bibinfo {author} {\bibfnamefont {A.}~\bibnamefont {Ross}},
  \bibinfo {author} {\bibfnamefont {T.}~\bibnamefont {Body}}, \bibinfo {author}
  {\bibfnamefont {M.}~\bibnamefont {Francisquez}}, \bibinfo {author}
  {\bibfnamefont {W.}~\bibnamefont {Zholobenko}}, \bibinfo {author}
  {\bibfnamefont {D.}~\bibnamefont {Coster}}, \bibinfo {author} {\bibfnamefont
  {O.}~\bibnamefont {Maj}}, \bibinfo {author} {\bibfnamefont {P.}~\bibnamefont
  {Manz}}, \bibinfo {author} {\bibfnamefont {F.}~\bibnamefont {Jenko}},
  \bibinfo {author} {\bibfnamefont {B.~N.}\ \bibnamefont {Rogers}},\ and\
  \bibinfo {author} {\bibfnamefont {K.~S.}\ \bibnamefont {Kang}},\ }\href
  {https://doi.org/10.1063/1.5089864} {\bibfield  {journal} {\bibinfo
  {journal} {Physics of Plasmas}\ }\textbf {\bibinfo {volume} {26}},\ \bibinfo
  {pages} {052517} (\bibinfo {year} {2019})}\BibitemShut {NoStop}%
\bibitem [{\citenamefont {Raissi}(2018)}]{Raissi_JMLR}%
  \BibitemOpen
  \bibfield  {author} {\bibinfo {author} {\bibfnamefont {M.}~\bibnamefont
  {Raissi}},\ }\href {http://jmlr.org/papers/v19/18-046.html} {\bibfield
  {journal} {\bibinfo  {journal} {Journal of Machine Learning Research}\
  }\textbf {\bibinfo {volume} {19}},\ \bibinfo {pages} {1} (\bibinfo {year}
  {2018})}\BibitemShut {NoStop}%
\bibitem [{\citenamefont {Vassilev}\ \emph {et~al.}(2020)\citenamefont
  {Vassilev}, \citenamefont {Efremov},\ and\ \citenamefont
  {Shadura}}]{AD_2020}%
  \BibitemOpen
  \bibfield  {author} {\bibinfo {author} {\bibfnamefont {V.}~\bibnamefont
  {Vassilev}}, \bibinfo {author} {\bibfnamefont {A.}~\bibnamefont {Efremov}},\
  and\ \bibinfo {author} {\bibfnamefont {O.}~\bibnamefont {Shadura}},\ }\href
  {https://doi.org/10.1051/epjconf/202024502015} {\bibfield  {journal}
  {\bibinfo  {journal} {EPJ Web Conf.}\ }\textbf {\bibinfo {volume} {245}},\
  \bibinfo {pages} {02015} (\bibinfo {year} {2020})}\BibitemShut {NoStop}%
\bibitem [{\citenamefont {Wang}\ \emph
  {et~al.}(2020{\natexlab{b}})\citenamefont {Wang}, \citenamefont {Wang},\ and\
  \citenamefont {Perdikaris}}]{wang2020eigenvector}%
  \BibitemOpen
  \bibfield  {author} {\bibinfo {author} {\bibfnamefont {S.}~\bibnamefont
  {Wang}}, \bibinfo {author} {\bibfnamefont {H.}~\bibnamefont {Wang}},\ and\
  \bibinfo {author} {\bibfnamefont {P.}~\bibnamefont {Perdikaris}},\
  }\href@noop {} {\bibinfo {title} {On the eigenvector bias of fourier feature
  networks: From regression to solving multi-scale pdes with physics-informed
  neural networks}} (\bibinfo {year} {2020}{\natexlab{b}}),\ \Eprint
  {https://arxiv.org/abs/2012.10047} {arXiv:2012.10047 [cs.LG]} \BibitemShut
  {NoStop}%
\bibitem [{\citenamefont {Cybenko}(1989)}]{cybenko_approximation_1989}%
  \BibitemOpen
  \bibfield  {author} {\bibinfo {author} {\bibfnamefont {G.}~\bibnamefont
  {Cybenko}},\ }\href@noop {} {\bibfield  {journal} {\bibinfo  {journal}
  {Mathematics of Control, Signals and Systems}\ }\textbf {\bibinfo {volume}
  {2}},\ \bibinfo {pages} {303} (\bibinfo {year} {1989})}\BibitemShut {NoStop}%
\bibitem [{\citenamefont {Zweben}\ \emph {et~al.}(2017)\citenamefont {Zweben},
  \citenamefont {Terry}, \citenamefont {Stotler},\ and\ \citenamefont
  {Maqueda}}]{Zweben_2017}%
  \BibitemOpen
  \bibfield  {author} {\bibinfo {author} {\bibfnamefont {S.~J.}\ \bibnamefont
  {Zweben}}, \bibinfo {author} {\bibfnamefont {J.~L.}\ \bibnamefont {Terry}},
  \bibinfo {author} {\bibfnamefont {D.~P.}\ \bibnamefont {Stotler}},\ and\
  \bibinfo {author} {\bibfnamefont {R.~J.}\ \bibnamefont {Maqueda}},\ }\href
  {https://doi.org/10.1063/1.4981873} {\bibfield  {journal} {\bibinfo
  {journal} {Review of Scientific Instruments}\ }\textbf {\bibinfo {volume}
  {88}},\ \bibinfo {pages} {041101} (\bibinfo {year} {2017})}\BibitemShut
  {NoStop}%
\bibitem [{\citenamefont {Glorot}\ and\ \citenamefont
  {Bengio}(2010)}]{GlorotAISTATS2010}%
  \BibitemOpen
  \bibfield  {author} {\bibinfo {author} {\bibfnamefont {X.}~\bibnamefont
  {Glorot}}\ and\ \bibinfo {author} {\bibfnamefont {Y.}~\bibnamefont
  {Bengio}},\ }in\ \href@noop {} {\emph {\bibinfo {booktitle} {JMLR W\&CP:
  Proceedings of the Thirteenth International Conference on Artificial
  Intelligence and Statistics}}},\ Vol.~\bibinfo {volume} {9}\ (\bibinfo {year}
  {2010})\ pp.\ \bibinfo {pages} {249--256}\BibitemShut {NoStop}%
\bibitem [{\citenamefont {Abadi}\ \emph {et~al.}(2015)\citenamefont {Abadi},
  \citenamefont {Agarwal}, \citenamefont {Barham}, \citenamefont {Brevdo},
  \citenamefont {Chen}, \citenamefont {Citro}, \citenamefont {Corrado},
  \citenamefont {Davis}, \citenamefont {Dean}, \citenamefont {Devin},
  \citenamefont {Ghemawat}, \citenamefont {Goodfellow}, \citenamefont {Harp},
  \citenamefont {Irving}, \citenamefont {Isard}, \citenamefont {Jia},
  \citenamefont {Jozefowicz}, \citenamefont {Kaiser}, \citenamefont {Kudlur},
  \citenamefont {Levenberg}, \citenamefont {Man\'{e}}, \citenamefont {Monga},
  \citenamefont {Moore}, \citenamefont {Murray}, \citenamefont {Olah},
  \citenamefont {Schuster}, \citenamefont {Shlens}, \citenamefont {Steiner},
  \citenamefont {Sutskever}, \citenamefont {Talwar}, \citenamefont {Tucker},
  \citenamefont {Vanhoucke}, \citenamefont {Vasudevan}, \citenamefont
  {Vi\'{e}gas}, \citenamefont {Vinyals}, \citenamefont {Warden}, \citenamefont
  {Wattenberg}, \citenamefont {Wicke}, \citenamefont {Yu},\ and\ \citenamefont
  {Zheng}}]{tensorflow2015-whitepaper}%
  \BibitemOpen
  \bibfield  {author} {\bibinfo {author} {\bibfnamefont {M.}~\bibnamefont
  {Abadi}}, \bibinfo {author} {\bibfnamefont {A.}~\bibnamefont {Agarwal}},
  \bibinfo {author} {\bibfnamefont {P.}~\bibnamefont {Barham}}, \bibinfo
  {author} {\bibfnamefont {E.}~\bibnamefont {Brevdo}}, \bibinfo {author}
  {\bibfnamefont {Z.}~\bibnamefont {Chen}}, \bibinfo {author} {\bibfnamefont
  {C.}~\bibnamefont {Citro}}, \bibinfo {author} {\bibfnamefont {G.~S.}\
  \bibnamefont {Corrado}}, \bibinfo {author} {\bibfnamefont {A.}~\bibnamefont
  {Davis}}, \bibinfo {author} {\bibfnamefont {J.}~\bibnamefont {Dean}},
  \bibinfo {author} {\bibfnamefont {M.}~\bibnamefont {Devin}}, \bibinfo
  {author} {\bibfnamefont {S.}~\bibnamefont {Ghemawat}}, \bibinfo {author}
  {\bibfnamefont {I.}~\bibnamefont {Goodfellow}}, \bibinfo {author}
  {\bibfnamefont {A.}~\bibnamefont {Harp}}, \bibinfo {author} {\bibfnamefont
  {G.}~\bibnamefont {Irving}}, \bibinfo {author} {\bibfnamefont
  {M.}~\bibnamefont {Isard}}, \bibinfo {author} {\bibfnamefont
  {Y.}~\bibnamefont {Jia}}, \bibinfo {author} {\bibfnamefont {R.}~\bibnamefont
  {Jozefowicz}}, \bibinfo {author} {\bibfnamefont {L.}~\bibnamefont {Kaiser}},
  \bibinfo {author} {\bibfnamefont {M.}~\bibnamefont {Kudlur}}, \bibinfo
  {author} {\bibfnamefont {J.}~\bibnamefont {Levenberg}}, \bibinfo {author}
  {\bibfnamefont {D.}~\bibnamefont {Man\'{e}}}, \bibinfo {author}
  {\bibfnamefont {R.}~\bibnamefont {Monga}}, \bibinfo {author} {\bibfnamefont
  {S.}~\bibnamefont {Moore}}, \bibinfo {author} {\bibfnamefont
  {D.}~\bibnamefont {Murray}}, \bibinfo {author} {\bibfnamefont
  {C.}~\bibnamefont {Olah}}, \bibinfo {author} {\bibfnamefont {M.}~\bibnamefont
  {Schuster}}, \bibinfo {author} {\bibfnamefont {J.}~\bibnamefont {Shlens}},
  \bibinfo {author} {\bibfnamefont {B.}~\bibnamefont {Steiner}}, \bibinfo
  {author} {\bibfnamefont {I.}~\bibnamefont {Sutskever}}, \bibinfo {author}
  {\bibfnamefont {K.}~\bibnamefont {Talwar}}, \bibinfo {author} {\bibfnamefont
  {P.}~\bibnamefont {Tucker}}, \bibinfo {author} {\bibfnamefont
  {V.}~\bibnamefont {Vanhoucke}}, \bibinfo {author} {\bibfnamefont
  {V.}~\bibnamefont {Vasudevan}}, \bibinfo {author} {\bibfnamefont
  {F.}~\bibnamefont {Vi\'{e}gas}}, \bibinfo {author} {\bibfnamefont
  {O.}~\bibnamefont {Vinyals}}, \bibinfo {author} {\bibfnamefont
  {P.}~\bibnamefont {Warden}}, \bibinfo {author} {\bibfnamefont
  {M.}~\bibnamefont {Wattenberg}}, \bibinfo {author} {\bibfnamefont
  {M.}~\bibnamefont {Wicke}}, \bibinfo {author} {\bibfnamefont
  {Y.}~\bibnamefont {Yu}},\ and\ \bibinfo {author} {\bibfnamefont
  {X.}~\bibnamefont {Zheng}},\ }\href {https://www.tensorflow.org/} {\bibinfo
  {title} {{TensorFlow}: Large-scale machine learning on heterogeneous
  systems}} (\bibinfo {year} {2015}),\ \bibinfo {note} {software available from
  tensorflow.org}\BibitemShut {NoStop}%
\bibitem [{\citenamefont {Kingma}\ and\ \citenamefont
  {Ba}(2014)}]{kingma2014adam}%
  \BibitemOpen
  \bibfield  {author} {\bibinfo {author} {\bibfnamefont {D.~P.}\ \bibnamefont
  {Kingma}}\ and\ \bibinfo {author} {\bibfnamefont {J.}~\bibnamefont {Ba}},\
  }\href@noop {} {\bibinfo {title} {Adam: A method for stochastic
  optimization}} (\bibinfo {year} {2014}),\ \Eprint
  {https://arxiv.org/abs/1412.6980} {arXiv:1412.6980 [cs.LG]} \BibitemShut
  {NoStop}%
\bibitem [{\citenamefont {Liu}\ and\ \citenamefont
  {Nocedal}(1989)}]{10.5555/3112655.3112866}%
  \BibitemOpen
  \bibfield  {author} {\bibinfo {author} {\bibfnamefont {D.~C.}\ \bibnamefont
  {Liu}}\ and\ \bibinfo {author} {\bibfnamefont {J.}~\bibnamefont {Nocedal}},\
  }\href@noop {} {\bibfield  {journal} {\bibinfo  {journal} {Math. Program.}\
  }\textbf {\bibinfo {volume} {45}},\ \bibinfo {pages} {503–528} (\bibinfo
  {year} {1989})}\BibitemShut {NoStop}%
\bibitem [{\citenamefont {Poulsen}\ \emph {et~al.}(2020)\citenamefont
  {Poulsen}, \citenamefont {Rasmussen}, \citenamefont {Wiesenberger},\ and\
  \citenamefont {Naulin}}]{multi_species}%
  \BibitemOpen
  \bibfield  {author} {\bibinfo {author} {\bibfnamefont {A.}~\bibnamefont
  {Poulsen}}, \bibinfo {author} {\bibfnamefont {J.~J.}\ \bibnamefont
  {Rasmussen}}, \bibinfo {author} {\bibfnamefont {M.}~\bibnamefont
  {Wiesenberger}},\ and\ \bibinfo {author} {\bibfnamefont {V.}~\bibnamefont
  {Naulin}},\ }\href {https://doi.org/10.1063/1.5140522} {\bibfield  {journal}
  {\bibinfo  {journal} {Physics of Plasmas}\ }\textbf {\bibinfo {volume}
  {27}},\ \bibinfo {pages} {032305} (\bibinfo {year} {2020})}\BibitemShut
  {NoStop}%
\bibitem [{\citenamefont {Mathews}\ \emph
  {et~al.}(2021{\natexlab{b}})\citenamefont {Mathews}, \citenamefont {Mandell},
  \citenamefont {Francisquez}, \citenamefont {Hughes},\ and\ \citenamefont
  {Hatch}}]{Mathews2021_FluidvGk}%
  \BibitemOpen
  \bibfield  {author} {\bibinfo {author} {\bibfnamefont {A.}~\bibnamefont
  {Mathews}}, \bibinfo {author} {\bibfnamefont {N.}~\bibnamefont {Mandell}},
  \bibinfo {author} {\bibfnamefont {M.}~\bibnamefont {Francisquez}}, \bibinfo
  {author} {\bibfnamefont {J.~W.}\ \bibnamefont {Hughes}},\ and\ \bibinfo
  {author} {\bibfnamefont {D.~R.}\ \bibnamefont {Hatch}},\ }\href@noop {}
  {\bibinfo {title} {Turbulent field fluctuations in gyrokinetic and fluid
  plasmas}} (\bibinfo {year} {2021}{\natexlab{b}}),\ \bibinfo {note} {in
  preparation}\BibitemShut {NoStop}%
\bibitem [{\citenamefont {Callen}(2006)}]{Callen_adiabatic}%
  \BibitemOpen
  \bibfield  {author} {\bibinfo {author} {\bibfnamefont {J.~D.}\ \bibnamefont
  {Callen}},\ }\href {http://homepages.cae.wisc.edu/{~}callen/book.html
  http://homepages.cae.wisc.edu/(~)callen/chap1.pdf} {\emph {\bibinfo {title}
  {Fundamentals of Plasma Physics}}},\ \bibinfo {edition} {draft}\ ed.\
  (\bibinfo  {publisher} {University of Wisconsin-Madison},\ \bibinfo {address}
  {Madison, Wisconsin},\ \bibinfo {year} {2006})\ Chap.\ \bibinfo {chapter}
  {Collective Plasma Phenomena}\BibitemShut {NoStop}%
\bibitem [{\citenamefont {Viezzer}\ \emph {et~al.}(2013)\citenamefont
  {Viezzer}, \citenamefont {P{\"u}tterich}, \citenamefont {Angioni},
  \citenamefont {Bergmann}, \citenamefont {Dux}, \citenamefont {Fable},
  \citenamefont {McDermott}, \citenamefont {Stroth},\ and\ \citenamefont
  {and}}]{Viezzer_2013}%
  \BibitemOpen
  \bibfield  {author} {\bibinfo {author} {\bibfnamefont {E.}~\bibnamefont
  {Viezzer}}, \bibinfo {author} {\bibfnamefont {T.}~\bibnamefont
  {P{\"u}tterich}}, \bibinfo {author} {\bibfnamefont {C.}~\bibnamefont
  {Angioni}}, \bibinfo {author} {\bibfnamefont {A.}~\bibnamefont {Bergmann}},
  \bibinfo {author} {\bibfnamefont {R.}~\bibnamefont {Dux}}, \bibinfo {author}
  {\bibfnamefont {E.}~\bibnamefont {Fable}}, \bibinfo {author} {\bibfnamefont
  {R.}~\bibnamefont {McDermott}}, \bibinfo {author} {\bibfnamefont
  {U.}~\bibnamefont {Stroth}},\ and\ \bibinfo {author} {\bibfnamefont {E.~W.}\
  \bibnamefont {and}},\ }\href {https://doi.org/10.1088/0029-5515/54/1/012003}
  {\bibfield  {journal} {\bibinfo  {journal} {Nuclear Fusion}\ }\textbf
  {\bibinfo {volume} {54}},\ \bibinfo {pages} {012003} (\bibinfo {year}
  {2013})}\BibitemShut {NoStop}%
\bibitem [{\citenamefont {Wersal}\ and\ \citenamefont
  {Ricci}(2015)}]{Wersal_2015}%
  \BibitemOpen
  \bibfield  {author} {\bibinfo {author} {\bibfnamefont {C.}~\bibnamefont
  {Wersal}}\ and\ \bibinfo {author} {\bibfnamefont {P.}~\bibnamefont {Ricci}},\
  }\href {https://doi.org/10.1088/0029-5515/55/12/123014} {\bibfield  {journal}
  {\bibinfo  {journal} {Nuclear Fusion}\ }\textbf {\bibinfo {volume} {55}},\
  \bibinfo {pages} {123014} (\bibinfo {year} {2015})}\BibitemShut {NoStop}%
\bibitem [{\citenamefont {Simakov}\ and\ \citenamefont
  {Catto}(2006)}]{Catto-paper}%
  \BibitemOpen
  \bibfield  {author} {\bibinfo {author} {\bibfnamefont {A.~N.}\ \bibnamefont
  {Simakov}}\ and\ \bibinfo {author} {\bibfnamefont {P.~J.}\ \bibnamefont
  {Catto}},\ }\href {https://doi.org/10.1063/1.2404555} {\bibfield  {journal}
  {\bibinfo  {journal} {AIP Conference Proceedings}\ }\textbf {\bibinfo
  {volume} {871}},\ \bibinfo {pages} {238} (\bibinfo {year}
  {2006})}\BibitemShut {NoStop}%
\bibitem [{\citenamefont {Huba}(2013)}]{Huba2013}%
  \BibitemOpen
  \bibfield  {author} {\bibinfo {author} {\bibfnamefont {J.~D.}\ \bibnamefont
  {Huba}},\ }\href {http://wwwppd.nrl.navy.mil/nrlformulary/} {\emph {\bibinfo
  {title} {Plasma Physics}}}\ (\bibinfo  {publisher} {Naval Research
  Laboratory},\ \bibinfo {address} {Washington, DC},\ \bibinfo {year} {2013})\
  pp.\ \bibinfo {pages} {1--71}\BibitemShut {NoStop}%
\end{thebibliography}%
\end{document}
%